\documentclass[11pt]{article}
\pdfoutput = 1
\usepackage[utf8]{inputenc}

\usepackage[top=1in,bottom=1in,left=1in,right=1in]{geometry}
\makeatletter \g@addto@macro\@floatboxreset\centering \makeatother
\usepackage{color}
\definecolor{darkgreen}{rgb}{0,0.5,0}
\definecolor{darkblue}{rgb}{0,0,0.6}
\definecolor{purple}{rgb}{0.4,.2,0.7}
\usepackage[noEucal]{main}
\usepackage{aas_macros,empheq,fancybox,graphicx,multirow}
\usepackage{aas_macros}
\usepackage{esint}
\usepackage{appendix}
\usepackage{mathabx}

\def\d{{\delta}}
\def\G{{\Gamma}}
\def\l{{\lambda}}
\def\S{{\Sigma}}
\def\o{{\omega}}
\def\ve{{\varepsilon}}
\def\p{{\partial}}
\def\s{{\sigma}}
\def\w{{\wedge}}

\def\ym{{\scriptscriptstyle YM}}

\def\dual{{\text{dual}}}

\usetikzlibrary{patterns}
\tikzstyle{edges}=[very thick]
\tikzstyle{particles}=[very thick, black]
\tikzstyle{interaction}=[circle, draw=black, pattern=nelines, very thick, inner sep=0pt, minimum size = 1.5cm]
\tikzstyle{positrons}=[black, very thick]
\colorlet{photons}{yellow!90!red!8!white}
\colorlet{phcont}{yellow!90!red}
\tikzstyle{int}=[circle, draw=black, fill=lightgray, very thick, inner sep=0pt, minimum size = 1.5cm]
\tikzstyle{hph}=[very thick, snake=snake, color=yellow]

\tikzset{
  on each segment/.style={
    decorate,
    decoration={
      show path construction,
      moveto code={},
      lineto code={
        \path [#1]
        (\tikzinputsegmentfirst) -- (\tikzinputsegmentlast);
      },
      curveto code={
        \path [#1] (\tikzinputsegmentfirst)
        .. controls
        (\tikzinputsegmentsupporta) and (\tikzinputsegmentsupportb)
        ..
        (\tikzinputsegmentlast);
      },
      closepath code={
        \path [#1]
        (\tikzinputsegmentfirst) -- (\tikzinputsegmentlast);
      },
    },
  },
  mid arrow/.style={postaction={decorate,decoration={
        markings,
        mark=at position .5 with {\arrow[#1]{stealth}}
      }}},
}

\tikzset{snake it/.style={decorate, decoration=snake}}
\tikzset{slope/.store in=\slope}
\pgfdeclarelayer{background}
\pgfdeclarelayer{foreground}
\pgfsetlayers{background,foreground}
\newcommand{\theslope}{0.7}
\pgfdeclarepatternformonly{nelines}
{\pgfpoint{-.2mm/\theslope}{-.2mm}}{\pgfpoint{2.2mm/\theslope}{2.2mm}}
{\pgfpoint{2mm/\theslope}{2mm}}
{
    \pgfsetlinewidth{1pt}
    \pgfpathmoveto{\pgfpoint{-.2mm/\theslope}{-.2mm}}
    \pgfpathlineto{\pgfpoint{2.2mm/\theslope}{2.2mm}}
    \pgfusepath{stroke}
}

\begin{document}
\thispagestyle{empty}

\begin{center}
    ~
    \vskip10mm

     {\LARGE  {\textsc{Shadows and Soft Exchange in Celestial CFT}}}
    \vskip10mm
    
    Daniel Kapec$^a$ and  Prahar Mitra$^{a,b}$ \\
    \vskip1em
    {\it
        $^a$ School of Natural Sciences, Institute for Advanced Study, Princeton, NJ 08540, USA\\ \vskip1mm
        $^b$ DAMTP, University of Cambridge, Wilberforce Road, Cambridge CB3 0WA, UK\\ \vskip1mm
         \vskip1mm
    }
    \vskip5mm
    \tt{danielkapec@fas.harvard.edu, pmitra@damtp.cam.ac.uk}
\end{center}
\vspace{10mm}

\begin{abstract}
\noindent
We study exponentiated soft exchange in $d+2$ dimensional gauge and gravitational theories using the celestial CFT formalism. These models exhibit spontaneously broken asymptotic symmetries generated by gauge transformations with non-compact support, and the effective dynamics of the associated Goldstone ``edge'' mode is expected to be $d$-dimensional. The introduction of an infrared regulator also explicitly breaks these symmetries so the edge mode in the regulated theory is really a $d$-dimensional pseudo-Goldstone boson. Symmetry considerations determine the leading terms in the effective action, whose coefficients are controlled by the infrared cutoff. Computations in this model reproduce the abelian infrared divergences in $d=2$, and capture the re-summed (infrared finite) soft exchange in higher dimensions. The model also reproduces the leading soft theorems in gauge and gravitational theories in all dimensions. Interestingly, we find that it is the shadow transform of the Goldstone mode that has local $d$-dimensional dynamics: the effective action expressed in terms of the Goldstone mode is non-local for $d>2$. We also introduce and discuss new magnetic soft theorems. Our analysis demonstrates that symmetry principles suffice to calculate soft exchange in gauge theory and gravity.

\end{abstract}
\pagebreak


{\hypersetup{linkcolor=black}
\small
\tableofcontents
}

\section{Introduction}
Gauge theories and gravitational theories in asymptotically flat spacetimes exhibit well-known universal infrared behavior. Soft theorems govern the form of scattering amplitudes at the boundaries of kinematic space, and repeated soft exchange exponentiates in a way that is independent of the microscopic details of the scattering process. Universality follows from symmetry, and both of these phenomena ultimately derive from the asymptotic symmetry structure of gauge theories in flat spacetime. 

In the Coulomb phase, gauge transformations with non-compact support act non-trivially on physical states. These ``large gauge transformations'' are broken by the perturbative vacuum, and correspond to flat directions in field space protected by symmetry. The unbroken global part of the gauge group gives rise to global charge conservation, but the local transformations are more akin to Goldstone bosons for an infinite-dimensional symmetry group. These universal features find natural expressions in the celestial CFT formalism. The isomorphism between the Lorentz group in $d+2$ dimensions and the Euclidean conformal group in $d$ dimensions can be exploited to recast standard $S$-matrix elements as ``celestial correlators'' that resemble those of a CFT$_d$. In this language, the unbroken symmetry group ensures the existence of a conserved current which is the shadow transform of a soft operator \cite{Kapec:2017gsg} and which satisfies the local Ward identities required of a CFT$_d$. In this paper, we demonstrate that the shadow transform of the Goldstone mode (corresponding to the broken symmetries) also has local $d-$dimensional dynamics, and that those dynamics fully reproduce the effects of soft exchange. Given that soft exchange occurs over large scales in the bulk, it is somewhat surprising that it can be captured by a local action in a lower dimension. We believe that our result lends support to the holographic interpretation of the celestial CFT formalism. There is a small body of previous work addressing soft exchange in celestial CFT, primarily in four dimensions \cite{Nande:2017dba,Himwich:2020rro,Arkani-Hamed:2020gyp,Magnea:2021fvy,Gonzalez:2021dxw,Nguyen:2021qkt,Kalyanapuram:2021tnl}. The methods and rationale in these treatments differ significantly from ours but the results for the case $d=2$ are equivalent.

Our analysis relies on the well known (dimension-independent) exponentiation of soft exchange in abelian gauge theory and gravity. If $\avg{ \CO_1 \cdots \CO_n }_\mu$ denotes the $S$-matrix element calculated with an infrared (IR) cutoff $\mu$ on loop-momenta and $\Lambda$ is the scale chosen to separate hard from soft, then 
\begin{equation}\notag
\avg{ \CO_1 \cdots \CO_n }_\mu  = e^{ - \G(\mu,\Lambda) } \avg{ \CO_1 \cdots \CO_n }_\L \; . 
\end{equation}
The quantity $\Gamma$ can be calculated from one-loop exchange diagrams and depends on the two scales $\mu$, $\Lambda$ and the quantum numbers of the external hard particles. The hard amplitude $\avg{ \CO_1 \cdots \CO_n }_\L$ is model dependent, but the exponential factor $e^{ - \G }$ is universal and one would like to compute it using symmetry principles. To do this, we separate fields into hard and soft pieces, integrate out the hard modes, and consider the effective action for the slow modes:
\begin{equation}\notag
\begin{split}
\avg{ \CO_1 \cdots \CO_n }_\mu &= \int [d \varphi_{s}  ] \int [ d \varphi_{h} ] e^{ i S_{\text{bulk}} [ \varphi_s , \varphi_h ] } \CO_1 \cdots \CO_n \\
&= \avg{ \CO_1 \cdots \CO_n }_\L \int  [d \varphi_{s} ] e^{ - S_{\text{soft}} [ \varphi_s ] - S_{\text{int}} [ \varphi_s , j ]  }  \;  .
\end{split}
\end{equation}
The form of the second equality relies on the identification of the soft modes $\varphi_s$ with the Goldstone edge modes associated to asymptotic symmetries. These degrees of freedom are effectively $d$-dimensional, and we expect their ``dynamics'' to be Euclidean. At this stage of the calculation, the external particles are accounted for by classical sources $j$  for $\varphi_s$ and  the matching condition reads
\begin{equation} \notag
\begin{split}
 \int  [d \varphi_{s} ] e^{ - S_{\text{soft}} [ \varphi_s ] - S_{\text{int}} [ \varphi_s , j ]  } = e^{ - \G}  \; .
\end{split}
\end{equation}
The form of the interaction term $S_{\text{int}} [ \varphi_s , j ]$ is completely fixed by the symmetry transformation properties of the external states under the broken symmetries. In order to determine $S_{\text{soft}} [ \varphi_s ]$ it is essential to understand the pattern of symmetry breaking in the IR regulated theory. The models that we consider exhibit infinite-dimensional groups of spontaneously broken symmetries $\mathcal{G}$ whose unbroken finite-dimensional subgroups we denote by $G$. For instance, in the case of $U(1)$ gauge theory on $\mathbb{R}^{1,d+1}$, the group  $\mathcal{G}$ consists of maps $S^d \to U(1)$ and $G=U(1)$ corresponds to the constant maps. If the $\mathcal{G}/G$ edge modes were true Goldstones, then the effective action  $S_{\text{soft}} [ \varphi_s ]$ would be invariant under a non-linear realization of the full infinite-dimensional group of $\mathcal{G}$ transformations. In the $U(1)$ example, a simple model for such an action is a $d$-dimensional $U(1)$ gauge theory in which one only integrates over (and does not quotient by) the trivial flat connections $C_a=\partial_aC(x)$. However, the introduction of the infrared regulator breaks these symmetries, so the effective action can also contain symmetry breaking terms that give an action to the Goldstone modes and $S_{\text{soft}} [ \varphi_s ]$ is the leading symmetry breaking term. It takes the form of a (non-local) mass term for the flat connection $C_a$ in the case of abelian gauge theory. 
In the models that we consider the resulting path integral is Gaussian and can be evaluated explicitly. The coefficient of $S_{\text{soft}} [ \varphi_s ]$ is then determined by matching onto the infrared divergence and it depends explicitly on the infrared cutoff. The resulting action can be used to compute the resummed soft exchange in any dimension. Interestingly, it is the shadow mode that has local dynamics: the action expressed in terms of the Goldstone variable is non-local when $d>2$. ``Integrating in'' a second degree of freedom related to external soft insertions results in a model that also reproduces the leading soft theorems in gauge and gravitational theories in all dimensions.

Although we do not treat the non-abelian case completely, our symmetry considerations do lead to a natural conjecture for the infrared action that fits in well with known analogies between non-linear sigma models in 2D and non-abelian gauge theories in 4D. In two dimensions the fluctuations of the non-abelian Goldstone bosons are strongly coupled and tend to disorder the system and restore the symmetry. A similar statement applies to the four-dimensional soft gluon when viewed as the order parameter for a spontaneously broken non-abelian large gauge symmetry: the soft gluon is strongly coupled in four and fewer dimensions, and the strong dynamics generically leads to the confining (unbroken) phase of gauge theory. In fact, a motivating point for this work is the observation that the infrared phases of Goldstone bosons in $d$ dimensions match those of gauge theories in $d+2$ dimensions. Both are infrared free (in the broken phase) when $d>2$ and generically strongly coupled when $d\leq2$. We hope to return to this case in future work and to continue to explore what is two-dimensional about four-dimensional gauge theory.

The outline of this paper is as follows. In section \ref{sec:Prelim}, we review soft theorems and soft exchange in gauge and gravitational theories in the language of conformal correlators. In section \ref{sec:MagSoft}, we extend these results to include magnetic and dyonic states, and introduce a variety of new magnetic soft theorems. In section \ref{sec:Actions} we use symmetry principles to derive the $d$-dimensional actions that reproduce soft exchange and soft theorems in $(d+2)$-dimensional models. Section \ref{sec:Conclusion} concludes with open questions, and appendix \ref{app:wardidderivation} contains derivations of important integral formulas.

\section{Preliminaries} \label{sec:Prelim}

In this section we review known results on the infrared sector of gauge theory and gravity, beginning with a discussion of the relationship between $S$-matrix elements and conformal correlation functions. We then review soft theorems and infrared divergences in the language of these correlators.

\subsection{$S$-matrix as a Conformal Correlator}

The Lorentz group in $d+2$ dimensions is isomorphic to the Euclidean conformal group in $d$ dimensions. This correspondence can be exploited to recast scattering amplitudes on $\mathbb{R}^{1,d+1}$ in terms of the correlation functions of a $d$-dimensional conformal field theory (CFT$_d$).

\subsubsection*{Poincar\'e and Conformal Algebra}
The Poincar\'e algebra is generated by the operators $P_\mu$ and $M_{\mu\nu}$  ($\mu,\nu=0,\cdots,d+1$) satisfying the commutation relations
\begin{equation}
\begin{split}\label{poincarealgebra}
[ P_\mu , P_\nu ] &= 0 \;, \qquad [ M_{\mu\nu} , P_\rho ] = 2 i \eta_{\rho[\mu} P_{\nu]} \;, \qquad [ M_{\mu\nu} , M_{\rho\s} ] = 4 i  \eta_{[\underline\rho[\mu} M_{\nu]\underline\s]} \; ,  
\end{split}
\end{equation}
where $\eta=\diag(-1,+1,\cdots,+1)$. The Lorentz generators $M_{\mu\nu}$ are linear combinations of the conformal generators $J_{ab}$, $D$, $T_a$ and $K_a$ ($a,b=1,\cdots,d$)
\begin{equation}
\begin{split}\label{confgen}
J_{ab} = M_{ab} \;, \qquad D = M_{d+1,0}  \;, \qquad T_a = M_{0,a} - M_{d+1,a} \;, \qquad K_a = M_{0,a} + M_{d+1,a}  \; . 
\end{split} 
\end{equation}
The Lorentz algebra implies that these generators satisfy the Euclidean conformal algebra
\begin{equation}
\begin{split}\label{confalgebra}
[  J_{ab} ,  J_{cd} ] &= 4 i \d_{[\underline c[a}  J_{b]\underline d]}  \;, \qquad [ J_{ab} ,  T_c ] = 2 i \d_{c[a}  T_{b]}  \;, \qquad [ J_{ab} ,  K_c ] = 2 i \d_{c[a}  K_{b]}  \;, \\
[ T_a , D  ] &= + i  T_a \;, \qquad [ K_a , D  ] = - i  K_a   \;, \qquad [ T_a ,  K_b ] = - 2 i ( \d_{ab}  D +  J_{ab} ) \;. \\
\end{split}
\end{equation}
The $J_{ab}$ generate $SO(d)$ rotations, $T_a$ and $K_a$ generate translations and special conformal transformations, and $D$ is the dilation operator. The mutually commuting generators $P^\mu$ decompose into two scalars $P^{\pm}=P^0 \pm P^{d+1}$ and a vector $P^a$ which mix under the action of the conformal algebra:
\begin{equation}
\begin{split}\label{eq:transAlgebra}
&[  J_{ab} ,  P^+ ] = 0  \;,  \qquad [ P^+ , D  ] = + i  P^+ \;, \qquad [ P^+ , T_a  ] = 0 \; , \qquad [ P^+ , K_a  ] = -2iP_a \; ,  \\
&[ J_{ab} ,  P^- ] =0  \;, \qquad [ P^- , D  ] = - i  P^- \;,  \qquad [ P^- , T_a  ] = -2iP_a \;,  \qquad [ P^- , K_a  ] = 0 \; , \\
&[ J_{ab} ,  P_c ] = 2 i \d_{c[a}  P_{b]}    \;,  \qquad [ P_a , D ] = 0 \;,  \qquad [ P_a , T_b ] = -i\delta_{ab}P^+ \; ,  \qquad [ P_a , K_b ] = -i\delta_{ab}P^- \;.  
\end{split}
\end{equation}

\subsubsection*{Momenta and Polarizations}

The $S$-matrix computes an overlap between multi-particle \emph{in} and \emph{out} states. It is a function of the momenta, spins and other quantum numbers associated to these states. In order to recast the $S$-matrix as a conformal correlator, it is useful to parameterize the momenta and polarization vectors in a way that makes the connection between the Lorentz group and the conformal group manifest. A convenient parameterization for the momenta is given by
\begin{equation}
\begin{split}\label{mompar}
p^\mu(\o,x) = \o {\hat p}^\mu ( \o , x ) \; , \qquad {\hat p}^\mu(\o,x) = {\hat q}^\mu(x) + ( m^2 / \o^2 )  n^\mu \; , 
\end{split}
\end{equation}
where ${\hat q}$ and $n$ are both null vectors given by
\begin{equation}\label{eq:nullMom}
\begin{split}
{\hat q}^\mu(x) = \frac{1}{2} ( 1 + x^2 , 2 x^a , 1 - x^2 ) \; , \qquad n^\mu = \frac{1}{2} ( 1 , 0^a , - 1 )  \; . 
\end{split}
\end{equation}
Momenta appearing in the scattering amplitude will be denoted by $p^\mu_i = p^\mu ( \o_i , x_i )$. We will adopt the convention in which outgoing momenta have $\eta_i = \text{sign}(\o_i) = +1$ and incoming momenta have $\eta_i = -1$. In these variables the Lorentz invariant inner product takes the form
\begin{equation}
\begin{split}
{\hat p}_i \cdot {\hat p}_j = - \frac{1}{2} [ x_{ij}^2 + m_i^2 / \o_i^2 + m_j^2 / \o_j^2  ] \; , \qquad x_{ij}^a = x_i^a - x_j^a \; .
\end{split}
\end{equation}

In addition to the momenta, spinning states are also labeled by their polarizations (little group representation). The $d$ independent polarization vectors for a massless spin one gauge boson will be taken to be
\begin{equation}\label{gluonpol}
\ve_a^\mu(x) \equiv \p_a \hat{q}^\mu(x) =  ( x_a , \delta_a^b, - x_a  ) \; . 
\end{equation}
These satisfy
\begin{equation}
\begin{split}
n \cdot \ve_a(x) = 0 \; , \qquad {\hat p}_i \cdot \ve_a(x_j) = ( x_{ij} )_a \; , \qquad \ve_a(x_i) \cdot \ve_b(x_j) = \d_{ab}  \; . 
\end{split}
\end{equation}
It is also useful to define the polarization sum
\begin{equation}
\begin{split}\label{polsum}
\Pi^{\mu\nu}(x) \equiv \ve^\mu_a(x) \ve^{a,\nu} (x) = \eta^{\mu\nu} + 2 n^{\mu} {\hat q}^{\nu}(x) + 2 n^{\nu} {\hat q}^{\mu}(x) \; . 
\end{split}
\end{equation}
The polarization tensor for a massless spin two graviton is given by
\begin{equation}
\begin{split}\label{gravpol}
\ve^{\mu\nu}_{ab}(x) \equiv\frac{1}{2}  [ \ve_a^{\mu}(x) \ve^{\nu}_b(x) + \ve_a^\nu(x) \ve^\mu_b(x)  ] - \frac{1}{d} \d_{ab} \Pi^{\mu\nu}(x)  \;, 
\end{split}
\end{equation}
and the corresponding polarization sum is 
\begin{equation}
\begin{split}\label{polsumgrav}
\Pi^{\mu\nu,\rho\s}(x) &\equiv  \ve^{\mu\nu}_{ab}(x) \ve^{ab,\rho\s}(x) = \frac{1}{2}\left[ \Pi^{\mu\rho} (x) \Pi^{\nu\s} (x) + \Pi^{\mu\s} (x) \Pi^{\nu \rho}(x) \right] - \frac{1}{d}\Pi^{\mu\nu} (x) \Pi^{\rho\s}(x) \; . 
\end{split}
\end{equation}
The Lorentz invariant phase space measure in this parameterization takes the form
\begin{equation}
\begin{split}
\int \frac{d^{d+1}p}{p^0} = \int  d^d x \int_0^\infty d\o \o^{d-1}  \;. 
\end{split}
\end{equation}
In section \ref{sec:IRDiv} we will discuss virtual soft exchange and off-shell loop momenta in loop integrals. A convenient parameterization is $\ell^\mu = \o [ {\hat q}^\mu (x) + \kappa n^\mu ]$. Integrals over off-shell momenta are given by
\begin{equation}
\begin{split}\label{momintegraloffshell}
\int d^{d+2} \ell = \frac{1}{2}  \int  d^d x  \int_{-\infty}^\infty d\o |\o|^{d+1}  \int_{-\infty}^\infty d\kappa \;. 
\end{split}
\end{equation}

\subsubsection*{Scattering States and Conformal Correlators}
\label{sec:in-out-op}

A scattering amplitude with $m$ outgoing particles and $n-m$ incoming particles is given by
\begin{equation}
\begin{split}\label{scat-amp}
\CA_n &=  \braket{ p_1 , \cdots , p_m }{ p_{m+1} , \cdots , p_n } \\
&= \bra{0} T \{ a_1^\text{out}(p_1) \cdots a_m^\text{out}(p_m) a_{m+1}^{\text{in}\dagger} (p_{m+1} ) \cdots a_{n}^{\text{in}\dagger} (p_n ) \} \ket{0} \; , 
\end{split}
\end{equation}
where $a_i^{\text{in}/\text{out}}(p_i)$ are the \emph{in} and \emph{out} annihilation operators. We will suppress the additional spin, color and/or flavor indices of these operators when they play no role. We can rewrite this amplitude in a suggestive way by defining
\begin{equation}
\begin{split}\label{Odef}
\CO_i(\o_i,x_i) \equiv a_i^\text{out}(p(\o_i,x_i)) \t(\o_i) + {\bar a}_i^{\text{in}\dagger}(-p(\o_i,x_i)) \t(-\o_i) \;. 
\end{split}
\end{equation}
According to this definition, when $\o_i < 0$ the operator inserts the CPT conjugate incoming particle. The scattering amplitude then takes the form 
\begin{equation}\label{eq:ConformalRep}
\begin{split}
\CA_n &=  \avg{ \CO_1(\o_1,x_1) \cdots \CO_n(\o_n,x_n) } \; . 
\end{split}
\end{equation}
Equation \eqref{eq:ConformalRep} is simply a rewriting of the scattering amplitude, and it is not immediately obvious why this representation is advantageous. Its utility derives from a reinterpretation of the Lorentz transformation properties of the operators \eqref{Odef}. Lorentz invariance fixes the transformation properties for an annihilation operator 
\begin{equation}
\begin{split}\label{LTa}
[ a_i^\text{out}(p) , M_{\mu\nu} ] = \CJ^{(i)}_{\mu\nu}   a_i^\text{out}(p) \; ,  \qquad   \CJ^{(i)}_{\mu\nu}  \equiv  \CL_{\mu\nu} + \CS^{(i)}_{\mu\nu} \;  .
\end{split}
\end{equation}
$\CL_{\mu\nu} = - 2 i p_{[\mu} \p_{p^{\nu]}}$ is the orbital angular momentum and $\CS^{(i)}_{\mu\nu}$ is the spin angular momentum operator for the $i$th particle. Combining  \eqref{LTa} with the definitions \eqref{confgen}, \eqref{Odef} and the parameterization \eqref{mompar}, one can determine the transformation properties of $\CO_i$ under the Euclidean conformal group. The transformations take a particularly simple form for massless operators (see \cite{Kapec:2017gsg} for details):
\begin{equation}
\begin{split}
\label{masslessLT}
[ \CO_i(\o,x) , T_a ] &=  i \p_a \CO_i(\o,x) \; , \\
[ \CO_i(\o,x) , J_{ab} ] &=  - i ( x_a  \p_b - x_b \p_a ) \CO_i(\o,x) + \CS^{(i)}_{ab}  \, \CO_i(\o,x) \; , \\
[ \CO_i(\o,x) , D ] &= i ( x^a \p_a - \o \p_\o ) \CO_i(\o,x) \;  , \\
[ \CO_i(\o,x) , K_a ] &= i  [ x^2  \p_a - 2 x_a x^b \p_b + 2 x_a \o \p_\o  ] \CO_i(\o,x) + 2 x^b \CS^{(i)}_{ab} \, \CO_i(\o,x) \;  . 
\end{split}
\end{equation}
Here, the $SO(d)$ little group generators $\CS^{(i)}_{ab}$ act in the representation determined by the spin of $\CO_i$. Equation \eqref{masslessLT} is precisely the transformation law for a $d$-dimensional conformal primary operator, with the important caveat that  $D$ is not diagonal in this basis.  Instead, the momentum eigenstate operators have a formal scaling dimension $\D = - \o \p_\o$. This is simply a reflection of the fact that momentum eigenstates are not simultaneously boost eigenstates. By performing a change of basis -- namely a Mellin transform with respect to $\o$ -- it is possible to define an equivalent set of operators which do diagonalize $D$,
\begin{equation}
\begin{split}\label{Mellin-transform}
\widehat{\CO}^\pm(\Delta,x) = \int_0^\infty d\o \o^{\D-1} \CO(\pm\o,x) \; , \qquad \D \in \frac{d}{2} + i \mrr \; . 
\end{split}
\end{equation}
The $\pm$ superscript distinguishes between \emph{in} and \emph{out} states. The restriction of $\D$ to the principal series is necessary in order for the operators ${\widehat \CO}$ to be normalizable. The inverse Mellin transform is used to map the operators back
\begin{equation}
\begin{split}\label{Inverse-Mellin-transform}
\CO(\pm|\o|,x)  = \int_{\frac{d}{2}-i\infty}^{\frac{d}{2}+i\infty} \frac{d\D}{2\pi i}  \, |\o|^{-\D} \widehat{\CO}^\pm(\Delta,x) \; . 
\end{split}
\end{equation}
Certain momentum eigenstate operators have universal behavior in the soft ($\o \to 0$) limit. Scattering amplitudes involving these special operators exhibit poles in this regime, and the residues can be isolated using a compact contour $\CC$ surrounding the origin in the complex $\omega$ plane:
\begin{equation}\label{compactcontour}
\widehat{\CO}(n,x) =\oint_\CC \frac{d\o}{2\pi i }  \,\o^{n-1}\CO(\o,x) \; . 
\end{equation}
It is not obvious whether this special class of operators is independent from those of the form \eqref{Mellin-transform}.

Operators corresponding to massive particles in momentum eigenstates also transform improperly (not as primaries) under special conformal transformations. For a massive scalar operator, the commutation relations are
\begin{equation}
\begin{split}\label{LorentzTransform}
[ \CO_i(\o,x) , T_a ] &=  i \p_a \CO_i(\o,x) \; , \\
[ \CO_i(\o,x) , J_{ab} ] &=  - i ( x_a  \p_b - x_b \p_a ) \CO_i(\o,x) \;  , \\
[ \CO_i(\o,x) , D ] &= i ( x^a \p_a - \o \p_\o ) \CO_i(\o,x)   \;  , \\
[ \CO_i(\o,x) , K_a ] &= i  [ ( x^2 + m_i^2/\o^2 ) \p_a - 2 x_a x^b \p_b + 2 x_a \o \p_\o ] \CO_i(\o,x) \;  .  \\
\end{split}
\end{equation}
Despite the more complicated form of this transformation law, there exists a basis transformation which maps the massive momentum eigenstate $\CO(\o,x)$ to a conformal primary operator
\begin{equation}
\begin{split}\label{massive-basis-transform}
\widehat{\CO}^\pm(\Delta,x) = m^{\D-d} \int d^d y \int_0^\infty d\o \o^{d-1} \CK_\D (   m / \o , y  ; x  )  \CO ( \pm \o , y ) \;  , \qquad \D \in \frac{d}{2} + i \mrr \; . 
\end{split}
\end{equation}
The kernel of this integral transform is the bulk-to-boundary propagator for a field of dimension $\Delta$ in Euclidean hyperbolic space $\mhh_{d+1}$
\begin{equation}
\begin{split}\label{bulk-bdy-prop}
\CK_\D (z , y ; x  )  = C_\D \left[ \frac{ z  }{  ( x - y )^2 + z^2  }  \right]^\D \; , \qquad C_\D = \frac{\G(\D)}{\pi^{\frac{d}{2}} \G ( \D - \frac{d}{2} ) } \; . 
\end{split}
\end{equation}
As in the massless case, the scaling dimensions are restricted to the principle series so that the operators are normalizable. To find the inverse transform, we use the split representation of the Dirac delta function in $\mhh_{d+1}$
\begin{equation}
\begin{split}
z^{d+1} \d ( z - z' ) \d^{(d)} ( x - x' )  = \frac{1}{2}\int_{\frac{d}{2}-i\infty}^{\frac{d}{2}+i\infty} \frac{d\D}{2\pi i}  \int d^d y \CK_\D  ( z , x ; y ) \CK_{d-\D} ( z' , x'  ; y )  \; . 
\end{split}
\end{equation}
Using this identity, the inverse transformation takes the form
\begin{equation}
\begin{split}\label{massive-inverse-basis-transform}
\CO(\pm|\o|,x)  &= \frac{1}{2} \int_{\frac{d}{2}-i\infty}^{\frac{d}{2}+i\infty} \frac{d\D}{2\pi i} \int d^d y m^{-\D}  \CK_{d-\D} (   m / |\o| ,x  ; y ) \widehat{\CO}^\pm(\Delta,y)  \; . 
\end{split}
\end{equation}
The operators $\widehat{\CO}^\pm(\Delta,x)$ varied over $\D$ are not independent of each other. Instead, they satisfy
\begin{equation}
\begin{split}
 \int d^d y \frac{C_{d-\D}}{ [ ( x - y )^2 ]^{d-\D}  } {\widehat \CO}^\pm(\Delta,y) = \widehat{\CO}^\pm(d-\D,x) \;  .
\end{split}
\end{equation}
This integral transform defines the shadow transform for scalar operators. The definition for more general operators is given in \eqref{shadow-def}.

The basis transformation for spinning massive states is more complicated since the massive little group is $SO(d+1)$ and its representations must be decomposed  into representations of $SO(d)$. We will not need the precise form of this generalized Mellin transform.

\subsubsection*{Summary}

The result of this section is that the Mellin-transformed $n$-point scattering amplitude on $\mathbb{R}^{1,d+1}$ 
\begin{equation}
\begin{split}\label{Andef}
\widehat{\CA}_n &=  \avg{ \widehat{\CO}^{\eta_1}_{1}(\Delta_1,x_1) \cdots \widehat{\CO}^{\eta_n}_{n}(\Delta_n, x_n) }  
\end{split}
\end{equation}
transforms as a $d$-dimensional conformal correlation function. The operators appearing on the RHS transform as conformal primaries: massless particles correspond to local operator insertions given by \eqref{Mellin-transform} while massive particles correspond to operator insertions given by \eqref{massive-basis-transform}.

The operators $\CO_i$ are generically labelled by additional spin, flavor, and/or color indices.  We only exhibit the spin indices when the particle is a gluon or graviton since they play a special role in what follows. The corresponding operators are denoted by $\CO_a$ and $\CO_{ab}$. These operators create one-particle states with polarization $\ve_a$ given by \eqref{gluonpol} or $\ve_{ab}$ given by \eqref{gravpol}, respectively.

\subsection{Soft Theorems and Conserved Currents}
\label{sec:softthmreview}

Soft theorems describe the universal behavior of scattering amplitudes involving one or more energetically soft (long wavelength) external particles. Long wavelength states are unable to resolve short-distance scattering processes, so amplitudes with soft insertions typically factorize into lower point amplitudes acted on by soft factors that are only sensitive to long distance data (like quantum numbers). This factorization is (semi-)universal and its form is independent of the microscopic details of the theory. Universal behavior is usually a consequence of an underlying symmetry, and work over the past several years has demonstrated that most of these universal formulas are reflections of the asymptotic symmetry structures  in gauge and gravitational theories in  flat space\cite{He:2014laa,He:2014cra,Kapec:2014opa,He:2015zea,Kapec:2016jld,He:2017fsb}. In this section, we review the soft photon, soft gluon and the leading soft graviton theorems in the language of conformal correlators.

\subsubsection*{Leading Soft Photon Theorem}

In any dimension, the soft photon theorem implies\footnote{This is only formally true in four dimensions, since both sides of \eqref{soft-photon-thm} generically vanish for scattering states with a finite number of photons due to the infrared divergence (see section \ref{sec:IRDiv}). In that case, the information content of the soft theorem resides in the form of the infrared finite dressed states.}
\begin{equation}
\begin{split}\label{soft-photon-thm}
\avg{ \CO_a ( \o , x ) \CO_1 \cdots \CO_n }_{C=0} ~~ \stackrel{\o\to0}{\longrightarrow} ~~ \frac{1}{\o} e \sum_i Q_i \frac{ {\hat p}_i \cdot \ve_a(x) }{ {\hat p}_i \cdot {\hat q}(x)  } \avg{ \CO_1 \cdots \CO_n }_{C=0}  \;  , 
\end{split}
\end{equation}
where $\CO_i \equiv \CO_i(\o_i,x_i)$ and $C = A |_{\p\ci}$ is the asymptotic connection. The subscript $C=0$ in \eqref{soft-photon-thm} indicates that the connection used to compare the electric charges at different points on the celestial sphere is taken to be trivial in standard quantum field theory (the electron has charge $-1$ no matter where it appears on the celestial sphere) so we will suppress this notation in what follows.\footnote{The role of this asymptotic connection is more fundamental in the non-abelian case, where it poses a potential obstruction to the definition of non-abelian dyonic charge.}

The soft theorem states that in the limit where the energy of the photon is small compared to the energies of the hard particles, the amplitude factorizes into a simple soft factor multiplied by a reduced scattering amplitude involving the remaining $n$ hard states. The gauge coupling constant is $e$ and $Q_i$ is the electric charge of the $i$th particle. In our conventions, the electric charge is defined by
\begin{equation}
\begin{split}
Q = \frac{1}{e^2}  \int_{S^2} \star F \;.
\end{split}
\end{equation}
We will assume that the gauge group is compact  ($U(1)$ as opposed to $\mrr$), so the charges are quantized
\begin{equation}
\begin{split}\label{el-chargequant}
Q_i = n_i \in \mzz \;  . 
\end{split}
\end{equation}
We define the leading ($n=1$) soft photon operator according to \eqref{compactcontour}
\begin{equation}\label{Sdef}
S_a(x) \equiv \frac{1}{e} \oint_\CC \frac{d\o}{2\pi i}  \CO_a(\o,x) \;  .
\end{equation}
 The contour integral isolates the residue of $\CO_a(\o,x)$ at $\o = 0$, and $S_a(x)$ has scaling dimension $\D=1$. Insertions of the soft operator are therefore fixed completely by the soft theorem 
\begin{equation}
\begin{split}\label{Sains}
\avg{ S_a(x) \CO_1 \cdots \CO_n } = \CJ_a (x) \avg{ \CO_1 \cdots \CO_n  } \;  ,  \qquad \CJ_a (x) = \p_a \sum_i Q_i \log [ - {\hat p}_i \cdot {\hat q}(x) ] \;  .
\end{split}
\end{equation}
Multiple insertions of this soft operator simply factorize
\begin{equation}
\begin{split}\label{mult-soft}
\avg{ S_{a_1}(y_1) \cdots S_{a_m}(y_m) \CO_1 \cdots \CO_n } = \CJ_{a_1}(y_1) \cdots \CJ_{a_m}(y_m) \avg{ \CO_1 \cdots \CO_n } \;  . 
\end{split}
\end{equation}
This formula is obtained by taking an $n+m$ point amplitude with $m$ external photons and then taking consecutive soft limits $\o_1,\cdots , \o_m \to 0$. Since the photon does not carry charge, the order of these limits is irrelevant.
Equations \eqref{Sains}-\eqref{mult-soft} together demonstrate that $S_a$ behaves as a flat abelian connection. 
In the absence of magnetic charges, the gauge field edge mode $C$ is flat as well 
\begin{equation}
\begin{split}
\label{eq:FlatAbelian}
d S = 0  \; , \qquad d C = 0 \quad \implies \quad S = d \phi \; , \qquad C = d \t \;. 
\end{split}
\end{equation}

It was demonstrated in \cite{Kapec:2017gsg} that the soft photon theorem implies the existence of a conserved $U(1)$ current in the celestial CFT$_d$. The definition of this operator involves the shadow transform, which maps a primary operator of weight $\D$ in the representation $\mathcal{R}$ of $SO(d)$ to a conformal primary operator of weight $d-\D$ and representation $\mathcal{R}$,
\begin{equation}
\begin{split}\label{shadow-def}
{\wt \CO}(x) \equiv \int d^d y \frac{ 1 }{ [ ( x - y )^2 ]^{d-\D} } \mathcal{R}  ( \CI(x-y) ) \cdot \CO(y) \; .
\end{split}
\end{equation}
The double shadow transform is proportional to the identity,
\begin{equation}
\begin{split}
{\wt {\wt \CO}} (x) = c_{\D,\mathcal{R}} \CO(x)\;  , \qquad \int d^d z \frac{ \mathcal{R} ( \CI(x-z) \CI(z-y) ) }{ [ ( x - z )^2 ]^\D [ ( z - y )^2 ]^{d-\D} } =  c_{\D,\mathcal{R}} \d^{(d)}(x - y )  \; . 
\end{split}
\end{equation}
For the spin $s$ representation, this coefficient is given by
\begin{equation}
\begin{split}\label{eq:norm}
c_{\D,s} = \frac{\pi^d (\D-1)(d-\D-1)\G(\frac{d}{2}-\D)\G(\D-\frac{d}{2})}{(\D-1+s)(d-\D-1+s)\G(\D)\G(d-\D)} \; . 
\end{split}
\end{equation}
The $U(1)$ conserved current is the shadow transform of the soft operator \eqref{Sdef}
\begin{equation}
\label{eq:CurrentShadow}
J_a(x) =  \frac{1}{2c_{1,1}} {\wt S}_a(x) \; .
\end{equation}
In $d>2$, $c_{1,1}$ is given by \eqref{eq:norm} while for $d=2$ direct computation gives $c_{1,1} = 4\pi^2$ (instead of $\pi^2$).\footnote{$c_{1,1}$ vanishes in odd dimensions, but the shadow integral in \eqref{eq:CurrentShadow} vanishes as well. A finite result is obtained by evaluating the RHS of \eqref{eq:CurrentShadow} for general $\D$ and then taking the $\D \to 1$ limit.} Note that $S_a(x)$ has dimension 1, so that its shadow has  $\Delta=d-1$ which is appropriate for a conserved current. Insertions of this current take the form
\begin{equation}
\begin{split}\label{Jains}
\avg{ J_a(x) \CO_1 \cdots \CO_n } = j_a (x) \avg{ \CO_1 \cdots \CO_n } \; , \qquad j_a (x) = \frac{1}{2c_{1,1}} {\wt \CJ}_a(x) \; , 
\end{split}
\end{equation}
and the divergence is given by (see appendix \ref{app:wardidderivation} for details)
\begin{equation}
\begin{split}
\label{eq:CurrentDiv}
\avg{ \p^a J_a(x) \CO_1 \cdots \CO_n  } =  \sum_i Q_i \CK_d(   m_i / \o_i , x_i ; x  )  \avg{  \CO_1 \cdots \CO_n } \;  .
\end{split}
\end{equation}
The Ward identity \eqref{eq:CurrentDiv} associates a non-local charge distribution to the massive charged states. Massless states have a localized charge distribution which can be seen using the property
\begin{equation}
\begin{split}\label{Kprop}
\lim_{z \to 0} \CK_d ( z , x ;  y  )  = \d^{(d)}(x-y) \; .
\end{split}
\end{equation}

\subsubsection*{Leading Soft Gluon Theorem}

The generalization of the soft photon theorem \eqref{soft-photon-thm} to non-abelian gauge theories is straightforward. The soft limit of a single gluon with polarization $a$ and color $I$ takes the form
\begin{equation}
\begin{split}\label{soft-gluon-thm}
\avg{ \CO^I_a ( \o , x ) \CO_1 \cdots \CO_n }_{C=0}  ~~ \stackrel{\o\to0}{\longrightarrow} ~~ \frac{1}{\o}  i g_\ym  \sum_i \frac{ {\hat p}_i \cdot \ve_a(x) }{ {\hat p}_i \cdot {\hat q}(x)  } T^I_i \avg{ \CO_1 \cdots \CO_n }_{C=0}  \; . 
\end{split}
\end{equation}
$T^I_i$ is a generator in the Lie algebra $\mfg$ (for the gauge group $G$) acting in the representation $R_i$ under which the $i$th particle transforms, and $g_{\ym}$ is the gauge coupling. The generators are normalized so that
\begin{equation}
\begin{split}
\big[ T^I , T^J \big] = f^{IJK} T^K \; , \qquad f^{IKL} f^{JKL} = \d^{IJ} \; , \qquad (T_{adj}^I)^{JK} = - f^{IJK}\; . 
\end{split}
\end{equation}
The notation $C=0$ signifies that the asymptotic connection used to compare color states at infinity is set to zero in conventional scattering amplitude calculations. This simplification is benign in the absence of chromo-magnetic flux. However, as we will discuss in Section \ref{MagSoftGluon}, this background connection is non-zero in the presence of chromo-dyons and significantly complicates the definition of the asymptotic symmetry group in non-abelian models. In this section we will only consider electrically charged states, so we suppress this notation. 

The leading soft gluon operator is defined as
\begin{equation}
\begin{split}\label{Sdef-ym}
S^I_a(x) \equiv \frac{1}{g_\ym} \oint_\CC \frac{d\o}{2\pi i}  \CO^I_a(\o,x) \; .
\end{split}
\end{equation}
Its correlation functions are completely controlled by the soft gluon theorem
\begin{equation}\label{Sains-ym}
\avg{ S^I_a(x) \CO_1 \cdots \CO_n } = \CJ^I_a(x) \avg{ \CO_1  \cdots  \CO_n  } \; ,  \qquad \CJ^I_a(x) = i  \p_a  \sum_i \log [ - {\hat p}_i \cdot {\hat q}(x)   ] \, T^I_i \; .
\end{equation}
Note that $\CJ_a^I(x)$ is a matrix which acts on the color indices of the amplitude $\CA_n$.

Multiple insertions of the operator $S_a^{I}$ are defined by taking an $n+m$ point amplitude with $m$ gluons and then taking the soft limits $\o_1,\cdots \o_m\to0$. However, since the gluon is itself charged under the gauge group, the order in which these limits are performed \emph{does} matter. We will adopt the prescription that the soft limits are taken in the ``left-to-right'' order
\begin{equation}
\begin{split}\label{multsoftdef}
& \avg{ S_{a_1}^{I_1}(y_1)  \cdots S_{a_m}^{I_m}(y_m)  \CO_1 \cdots \CO_n } \\
&\qquad \qquad \qquad \equiv \frac{1}{g^m_\ym}\oint_{\CC_m} \frac{d\o_m}{2\pi i}  \cdots  \oint_{\CC_1} \frac{d\o_1}{2\pi i} \avg{  \CO^{I_1}_{a_1}(\o_1,y_1)  \cdots  \CO^{I_m}_{a_m}(\o_m,y_m)  \CO_1 \cdots \CO_n }  \; ,
\end{split}
\end{equation}
where $\CC_1 \subset \CC_2 \subset \cdots \subset \CC_m$ are nested contours in the complex $\omega$ plane. The soft operator $S$ and gauge field edge mode $C$ satisfy a non-abelian flatness condition
\begin{equation}
\begin{split}\label{NAflatcond}
d S + C \w S + S \w C  = 0 \; , \quad   d C + C \w C  = 0 \quad \implies \quad S = d \phi + [ C , \phi ] \; , \quad C = U d U^{-1}  \; . 
\end{split}
\end{equation}
We can also construct a non-abelian conserved current 
\begin{equation}
\begin{split}\label{eq:YMCurrentShadow}
J_a^I(x) = \frac{1}{2c_{1,1}} {\wt S}^I_a (x) \; , \qquad \avg{ J_a^I(x) \CO_1 \cdots \CO_n } = j^I_a(x) \avg{ \CO_1 \cdots \CO_n } 
\end{split}
\end{equation}
where $j^I_a(x) = \frac{1}{2c_{1,1}} {\wt \CJ}^I_a (x)$. Its divergence is given by
\begin{equation}
\begin{split}\label{eq:CurrentDiv-ym}
\avg{ \p^a J^I_a(x)\CO_1 \cdots \CO_n    } = i  \sum_i \CK_d(  m_i / \o_i , x_i ; x  )  T^I_i \avg{ \CO_1 \cdots \CO_n  } \; .
\end{split}
\end{equation}

\subsubsection*{Leading Soft Graviton Theorem}

The leading soft graviton theorem was first derived by Weinberg \cite{Weinberg:1965nx}. The soft limit of a single graviton with polarization $\varepsilon_{ab}$ takes the form
\begin{equation}
\begin{split}\label{soft-grav-thm}
\avg{ \CO_{ab} ( \o , x ) \CO_1 \cdots \CO_n }_{C=0}  ~~ \stackrel{\o\to0}{\longrightarrow} ~~   \frac{\kappa}{2\omega} \sum_i  \frac{ p_{i\mu} p_{i\nu} \ve^{\mu\nu}_{ab} (q)  }{ p_i \cdot \hat{q} } \avg{  \CO_1 \cdots \CO_n }_{C=0}  \; . 
\end{split}
\end{equation}
Here $\kappa^2=32\pi G$ and $C_{ab} = r^{-1} h_{ab} |_{\p \ci}$ ($h_{\mu\nu}=g_{\mu\nu}-\eta_{\mu\nu}$ is the metric fluctuation). The subscript $C=0$ in \eqref{soft-grav-thm} describes the BMS frame in which the scattering amplitudes are evaluated and with respect to which the energy-momentum of the particles are defined. This is the gravitational analogue of the flat connection $C = 0$ used in gauge theory scattering amplitudes. We will suppress this notation in the rest of this section.

The leading soft graviton operator is defined according to \eqref{compactcontour}
\begin{equation}
\begin{split}
N_{ab}(x) \equiv \frac{2}{\kappa} \oint_\CC \frac{d\o}{2\pi i} \CO_{ab}(\o,x) \;  
\end{split}
\end{equation}
 and has scaling dimension $\D=1$. Insertions of this operator take the form
\begin{equation}
\avg{ N_{ab}(x) \CO_1 \cdots \CO_n }  =  \CJ_{ab}(x)\, \avg{ \CO_1 \cdots \CO_n } \; ,  
\end{equation}
where
\begin{equation}
\begin{split}\label{Jabdef}
\CJ_{ab}(x) &\equiv - \left( \p_a \p_b - \frac{1}{d} \d_{ab} \p^2 \right) \sum_i \o_i [ - {\hat p}_i \cdot {\hat q}(x) ]  \log [ - {\hat p}_i \cdot {\hat q}(x) ] \;  .\\
\end{split}
\end{equation}
Multiple insertions of $N_{ab}$ are calculated by taking simultaneous soft limits of a scattering amplitude involving multiple external gravitons. Although general relativity is non-linear, its infrared dynamics is effectively abelian since the charge of the theory is energy-momentum. The order of soft limits is therefore immaterial and one finds
\begin{equation}
\begin{split}
&\avg{N_{a_1b_1}(y_1) \cdots N_{a_mb_m}(y_m) \CO_1 \cdots \CO_n } = \CJ_{a_1 b_1} ( y_1 ) \cdots \CJ_{a_m b_m} ( y_m ) \avg{ \CO_1 \cdots \CO_n } \;  .
\end{split}
\end{equation}
From this formula, it follows that $N_{ab}$ satisfies the following ``flatness condition''
\begin{equation}
\begin{split}
\p_{[a} N_{b]c} (x) - \frac{1}{d-1} \d_{c[a} \p^d N_{b]d} (x)  = 0 \quad \implies \quad N_{ab}(x) = 2 \left( \p_a \p_b - \frac{1}{d} \d_{ab} \p^2 \right)N(x) \;  .
\end{split}
\end{equation}
A similar property is also satisfied by the gravitational connection $C_{ab}(x)$ which then implies
\begin{equation}
\begin{split}
C_{ab}(x) = 2 \left( \p_a \p_b - \frac{1}{d} \d_{ab} \p^2 \right) C(x) \; . 
\end{split}
\end{equation}
This condition is the higher-dimensional analogue of the four-dimensional Christodoulou-Klainerman constraint \cite{Strominger:2013jfa}.

The leading soft graviton theorem implies the existence of $d+2$ conserved currents $P^+_a(x)$, $P^b_a(x)$ and $P^-_a(x)$ whose charges are the translation generators in \eqref{eq:transAlgebra}. However, not all of these currents are independent. Rather, they sit in a single conformal multiplet that mixes under Lorentz (conformal) transformations. Up to improvement terms, \eqref{eq:transAlgebra} implies
\begin{equation}
\begin{split}
P^b_a(x)  = \frac{i}{2} [ P^+_a(x) , K^b ] \; , \qquad  P^-_a (x)  = \frac{i}{d} [ P^b_a(x) , K_b ] \; . 
\end{split}
\end{equation}
The current $P^+_a(x)$ can be constructed from the shadow transform of the soft graviton operator as
\begin{equation}
\begin{split}\label{eq:TranslationCurrent}
P^+_a(x) = \frac{1}{4 c_{1,2} } \p^b {\wt N}_{ab}(x) \; .
\end{split}
\end{equation}
In $d>2$, $c_{1,2}$ is given by \eqref{eq:norm} while for $d=2$ direct computation gives $c_{1,2} = \pi^2$ (instead of $\pi^2/4$). The soft operator $N_{ab}$ has dimension $1$ so the current has $\D = d$. The unusual dimension for this current arises because the conserved charge $P^+$ is itself dimensionful according to \eqref{eq:transAlgebra}. Using this definition, we find
\begin{equation}
\begin{split}
\avg{ P^+_a(x) \CO_1 \cdots \CO_n } = p_a^+(x) \avg{ \CO_1 \cdots \CO_n } \; , 
\end{split}
\end{equation}
where
\begin{equation}
\begin{split}\label{jabdef-gr}
p^+_a(x) =  \p^b j_{ab}(x) , \qquad j_{ab}(x) =  -\frac{1}{4c_{1,2}} {\wt \CJ}_{ab}(x) \;. 
\end{split}
\end{equation}
The divergence of this current is then given by (see appendix \ref{app:wardidderivation})
\begin{equation}
\begin{split}\label{eq:CurrentDivGR}
\avg{ \p^a P^+_a(x) \CO_1 \cdots \CO_n } =  \sum_i m_i \CK_{d+1} ( m_i/\o_i , x_i ; x ) \avg{  \CO_1 \cdots \CO_n } \; .
\end{split}
\end{equation}
As in the abelian case, the Ward identity \eqref{eq:CurrentDivGR} associates a non-local energy-momentum distribution to massive charged states. Massless states have a localized energy-momentum as can be seen from
\begin{equation}
\begin{split}
\lim_{z \to 0} z \,  \CK_{d+1} ( z , x ; y ) = \d^{(d)} ( x - y ) \; . 
\end{split}
\end{equation}

\subsection{Infrared Divergences in Abelian Theories}\label{sec:IRDiv}

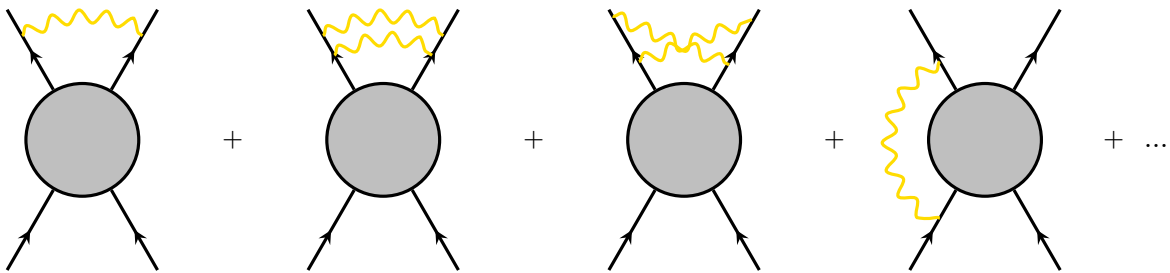
\begin{figure}[h!]
\begin{tikzpicture}
[node distance=2cm]
\begin{pgfonlayer}{foreground}
\node[int] (int1) at (0,0) {};
\node[draw=white] (plus1) [right of=int1] {$+$};
\node[int] (int2) [right of=plus1] {};
\node[draw=white] (plus2) [right of=int2] {$+$};
\node[int] (int3) [right of=plus2] {};
\node[draw=white] (plus3) [right of=int3] {$+$};
\node[int] (int4) [right of=plus3] {};
\node[draw=white] (plus4) [right of=int4] {$+ ~~...$};
\end{pgfonlayer}
\begin{pgfonlayer}{background}
\path [draw, positrons, postaction={on each segment={mid arrow=black}}] (int1) -- (120:2cm);
\path [draw, positrons, postaction={on each segment={mid arrow=black}}] (-60:2cm) -- (int1);
\path [draw, particles, postaction={on each segment={mid arrow=black}}] (int1) -- (60:2cm);
\path [draw, particles, postaction={on each segment={mid arrow=black}}] (-120:2cm) -- (int1);
\path[draw=phcont, snake it, very thick] (120:1.6cm) arc (120:60:1.6cm);
\path [draw, positrons, postaction={on each segment={mid arrow=black}}] (int2) -- +(120:2cm);
\path [draw, positrons, postaction={on each segment={mid arrow=black}}] ($(int2) +(-60:2cm)$) -- (int2);
\path [draw, particles, postaction={on each segment={mid arrow=black}}] (int2) -- +(60:2cm);
\path [draw, particles, postaction={on each segment={mid arrow=black}}] ($(int2) +(-120:2cm)$) -- (int2);
\path[draw=phcont, snake it, very thick] (int2)+(120:1.6cm) arc (120:60:1.6cm);
\path[draw=phcont, snake it, very thick] (int2)+(120:1.3cm) arc (120:60:1.3cm);
\path [draw=blue, positrons, postaction={on each segment={mid arrow=black}}] (int3) -- +(120:2cm);
\path [draw=blue, positrons, postaction={on each segment={mid arrow=black}}] ($(int3) +(-60:2cm)$) -- (int3);
\path [draw, particles, postaction={on each segment={mid arrow=black}}] (int3) -- +(60:2cm);
\path [draw, particles, postaction={on each segment={mid arrow=black}}] ($(int3) +(-120:2cm)$) -- (int3);
\path[draw=phcont, snake it, very thick] (int3)+(120:1.9cm) --  (8.6,1);
\path[draw=phcont, snake it, very thick] (int3)+(120:1.2cm) -- (8.9,1.6);
\path [draw=black, positrons, postaction={on each segment={mid arrow=black}}] (int4) -- +(120:2cm);
\path [draw=black, positrons, postaction={on each segment={mid arrow=black}}] ($(int4) +(-60:2cm)$) -- (int4);
\path [draw, particles, postaction={on each segment={mid arrow=black}}] (int4) -- +(60:2cm);
\path [draw, particles, postaction={on each segment={mid arrow=black}}] ($(int4) +(-120:2cm)$) -- (int4);
\path[draw=phcont, snake it, very thick] (int4)+(120:1.2cm) arc (120:240:1.2cm);
\end{pgfonlayer}
\end{tikzpicture}
\caption{Some typical diagrams contributing to infrared divergences }\label{fig:loops}
\label{IRdiagrams}
\end{figure}

In any theory with long-range forces, the naive scattering amplitude defined in \eqref{scat-amp} vanishes in four dimensions due to infrared divergences. Formally, these infrared divergences arise due to low energy virtual particle exchange in diagrams like figure \ref{fig:loops}. Physically, the divergences signify that the charged asymptotic scattering states are not single particle states but rather coherent states of charged particles and bremsstrahlung. 
Although each diagram is separately divergent, the infinite sum exponentiates and the amplitude actually vanishes: $\CA_n \sim e^{-\infty} = 0$. There is simply zero probability to scatter into a state with a finite number of photons or gravitons in four dimensions: the power radiated in a generic scattering process does not vanish at zero frequency, and so infinitely many low energy particles are produced \cite{mott_1931,Bloch:1937pw}. In higher dimensions this is no longer the case and exclusive scattering states with finite numbers of massless quanta can be defined. The soft contributions depicted in figure \ref{fig:loops} can be calculated and exponentiated in any number of dimensions, but it is only in four dimensions that the contribution is divergent. 

Standard treatments of this subject restrict attention to inclusive observables defined by a trace over ``unobservable soft quanta'' in the initial and final states. Order by order in perturbation theory, the infinite volume of phase space available to soft external quanta cancels the vanishing probability to scatter into an exclusive state with a fixed number of photons. Although perfectly adequate for practical purposes, this approach to the problem conflates two separate issues and obscures the underlying physics. It is certainly the case that any conceivable experimental apparatus has a finite energy sensitivity. In quantum mechanics, when an observer lacks access to a particular component of the Hilbert space, he performs a partial trace in order to obtain a reduced density matrix which can be used to compute observables. In a realistic scenario, any collider observable in any number of dimensions should include a trace over the unobservable soft quanta. However, the ``unobservable'' scale relevant to the trace is set by the detector sensitivity and is not related to the infrared divergence. An inadequate higher-dimensional detector might require a trace over states with relatively short wavelengths which are certainly produced with nonzero probability. This inclusive calculation, while relevant to the experiment, is not related to any infrared divergence since the exclusive observable is well defined in higher dimensions.

Taking a partial trace generically turns a pure density matrix into a mixed state. In order to discuss fine-grained questions like unitarity or the information content of soft quanta, one needs access to the actual quantum mechanical transition amplitudes in Hilbert space. The difficulty in defining an $S$-matrix in four-dimensional asymptotically flat space highlights a deficiency in a particular formalism but does not preclude the calculation of exact transition probabilities. 
The key physical observation relevant to the definition of an infrared finite $S$-matrix was made by Chung, Kibble and Faddeev-Kulish \cite{Chung:1965zza,Kibble:1968sfb,Kibble:1969ip,Kibble:1969ep,Kibble:1969kd,Kulish:1970ut}. The asymptotic Hamiltonian, which determines the early and late time evolution of scattering states, is not a free Hamiltonian. In the charged sectors of the Hilbert space, this Hamiltonian has matrix elements connecting states with different numbers of photons. A stationary state invariant under early/late time evolution must therefore contain an infinite number of photons corresponding to bremsstrahlung clouds. Computations involving these dressed states are cumbersome and appear to be plagued with ambiguities. One cannot help but feel that the ultimate treatment has yet to be discovered, but there are indications that the celestial scattering amplitudes and the conformal basis contain the appropriate formalism to address the problem\cite{Kapec:2017tkm,Arkani-Hamed:2020gyp}. In this paper we focus on understanding the soft virtual exchange in exclusive amplitudes and leave the proper determination of the scattering states to future work.  

There are several equivalent methods for regulating and cancelling infrared divergences in scattering amplitudes (see Chapter 13 of Weinberg \cite{Weinberg:1995mt} for a thorough discussion of the abelian case). Each of these methods breaks or alters a symmetry of the problem. One common approach introduces a hard IR cutoff $\mu$ for the photon. Loop integrals over virtual momenta are to be performed only in the range $|\o| > \mu$. A second scale $\L$ is chosen in order to separate out the contribution of soft particles which are defined to have $\mu < |\o| < \L$. Provided that  $\mu$ and $\L$ are taken to be much smaller than the typical energy scale of the scattering amplitude, the contributions of these quanta can be computed exactly and exponentiated. Hard particles with $|\o| > \L$ contribute IR finite (but generally incalculable and non-exponentiating) corrections. The general structure of a scattering amplitude is then
\begin{equation}
\begin{split}\label{IRgen}
\avg{ \CO_1 \cdots \CO_n }_\mu  = e^{ - \G } \avg{ \CO_1 \cdots \CO_n }_\L \; . 
\end{split}
\end{equation}
Here $\avg{ \CO_1 \cdots \CO_n }_\mu $ denotes the full scattering amplitude with IR cutoff $\mu$ and loop integrals performed with $|\o| > \mu$, while $\avg{ \CO_1 \cdots \CO_n }_\L$ denotes the amplitude with loop integrals performed in the range $|\o| > \L$. The soft  contribution resides entirely in the exponential factor $e^{-\G}$ which depends on the scales $\mu$, $\L$ and on the quantum numbers of the hard states.

\subsubsection*{Abelian Gauge Theories}
\label{sec:IRdivabelian}

In an abelian gauge theory, the infrared divergent piece of the full scattering amplitude is the exponentiation of the single-exchange contribution shown in the first diagram of figure \ref{IRdiagrams}. This dramatic simplification occurs because soft photons do not emit other soft particles and fails in non-abelian models (and in models with massless charged matter). The diagram is calculated using the soft photon theorem. Summing over the propagating internal states, one finds\footnote{The $i\e$ prescription for $i=j$ is a bit different and $- p_j \cdot \ell - i \e$ is replaced with $- p_i \cdot \ell + i \e$.\label{ieprescription}}
\begin{equation}
\begin{split}\label{ampfactor}
\G_{\text{ph}} =   \frac{i e^2}{2} \sum_{i,j} Q_i Q_j \int_\mu^\L \frac{d^{d+2} \ell}{(2\pi)^{d+2} } \frac{ p^\mu_i p^\nu_j \Pi_{\mu\nu}(\ell) }{ ( \ell^2 - i \e  ) ( p_i \cdot \ell - i \e ) ( - p_j \cdot \ell - i \e ) } \; .
\end{split}
\end{equation}
The integral is taken over the range $\mu < |\o | < \L$ and represents only the contribution of the ``soft quanta.'' Note that this piece of the integral can be separated out in any dimension even though it is not infrared divergent when $d>2$. The polarization sum $\Pi_{\mu\nu}(\ell)$ defined in \eqref{polsum} arises from the numerator of the photon propagator. This is of the form $\Pi_{\mu\nu}(\ell) = \eta_{\mu\nu} + \text{\emph{gauge-dependent terms}}$ so we could further simplify the expression by sending $\Pi_{\mu\nu}(\ell) \to \eta_{\mu\nu}$. 

This integral is evaluated explicitly in \cite{Weinberg:1995mt}. For our purposes it is  more useful to retain $\Gamma_{\text{ph}}$ as a $d$-dimensional integral. To do this, we parameterize all momenta using \eqref{mompar} and use \eqref{momintegraloffshell} to find
\begin{equation}
\begin{split}\label{Iijeval}
\G_{\text{ph}} &= - \frac{e^2}{4\pi} \sum_{i,j} Q_i Q_j  \int \frac{d^d x}{(2\pi)^d} \int\limits_{\mu<|\o|<\L} d\o |\o|^{d-3} \\
&\qquad \qquad \qquad \times \int_{-\infty}^\infty \frac{d\kappa}{2\pi i} \frac{  2 {\hat p}^\mu_i {\hat p}^\nu_j \Pi_{\mu\nu} (x) }{ ( \kappa + i \e  ) ( \kappa - 2 {\hat p}_i \cdot {\hat q}(x)  + i \eta \eta_i \e ) ( \kappa - 2 {\hat p}_j \cdot {\hat q}(x) - i \eta \eta_j \e ) }  \; . 
\end{split}
\end{equation}
We perform the integral over $\kappa$ via contour integration. When $\eta_i = \eta_j$ we close the contour in the upper half plane. When $\eta_i = - \eta_j = - \eta$, we close the contour in the lower half plane and we pick up the pole at $\kappa=-i\e$. Finally, when $\eta_i = - \eta_j = \eta$ we close the contour in the upper half plane. This contribution vanishes since there are no poles in the upper half plane. We can then perform the integral over $\o$ and we are left with a $d$-dimensional integral. We can also simplify further by setting $\Pi_{\mu\nu}(x) = \ve^a_\mu (x) \ve_{a\nu}(x)$. Plugging all of this into \eqref{ampfactor} and using the language of conformal correlators, we find
\begin{equation}
\begin{split}\label{soft-factorization}
\G_{\text{ph}} = \a  ( A_1 + 2\pi i A_2 ) \; ,  \qquad \a = \frac{e^2}{8\pi} \int_{\mu}^\L d\o \o^{d-3} \; ,
\end{split}
\end{equation}
and
\begin{equation}
\begin{split}\label{Adefabelian}
A_1 &=  \int  \frac{d^d x}{(2\pi)^d} [\CJ_a(x)]^2   \; , \qquad A_2 =  \int_{-\infty}^{\infty} \frac{d\nu}{2\pi}  \int  \frac{d^d x}{(2\pi)^d} [ | \CJ_a ^+(\nu,x) |^2   + | \CJ_a^-(\nu,x) |^2  ]_r \; .
\end{split}
\end{equation}
$\CJ_a(x)$ is defined in \eqref{Sains} and
\begin{equation}
\begin{split}\label{Japmdef}
\CJ_a^\pm (\nu,x) \equiv \p_a \sum_{i\in{\text{out} (+)\atop\text{in} (-)} } Q_i  \frac{ [ -  {\hat p}_i \cdot {\hat q}(x) ]^{i\nu}}{i\nu} \; , \qquad \CJ_a (x) = \lim_{\nu \to 0} [ \CJ_a^+(\nu,x) + \CJ_a^-(\nu,x) ] \; .
\end{split}
\end{equation}
The $[~]_r$ symbol removes any $i=j$ terms in the integrand.

In $d>2$, $\G_{\text{ph}} \to 0$ polynomially as $\L,\mu \to 0$ which shows that soft photons do not contribute any divergences to the scattering amplitude. On the other hand, in $d=2$, $\G_{\text{ph}} \to \infty$ logarithmically implying that the scattering amplitude itself vanishes as a power law as $\mu\to0$.

\subsubsection*{Gravitational Theories}

As in abelian gauge theory, the total graviton contribution to the infrared divergence is the exponentiation of the single virtual soft graviton contribution. This simplification holds because, although general relativity is strongly nonlinear, the charge of the theory is energy-momentum and long wavelength modes are weakly interacting. The infrared divergent phase in this case is 
\begin{equation}
\begin{split}\label{soft-factorization-gr}
\G_{\text{gr}} = \a_{\text{gr}} ( A_1^{\text{gr}}  + 2\pi i A_2^{\text{gr}}   ) \;  , \qquad \a_{\text{gr}} = \frac{\kappa^2}{32\pi} \int_{\mu}^\L d\o \o^{d-3} \; .
\end{split}
\end{equation}
The integrals in the exponent are
\begin{equation}
\begin{split}\label{ABdef}
A_1^{\text{gr}} &=  \int \frac{d^d x}{(2\pi)^d} [ \CJ_{ab} (x) ] ^2 \; , \qquad A_2^{\text{gr}} = \int_{-\infty}^\infty \frac{d\nu}{2\pi}  \int  \frac{d^d x}{(2\pi)^d}  [ |\CJ_{ab}^+(\nu,x) |^2 + |\CJ_{ab}^-(\nu,x) |^2 ]_r \; ,
\end{split}
\end{equation}
where $\CJ_{ab}(x)$ is defined in \eqref{Jabdef} and 
\begin{equation}
\begin{split}\label{Jabpmdef}
\CJ_{ab}^\pm (\nu,x) &= \left[ \p_a \p_b - \frac{1}{d} \d_{ab} \p^2 \right]    \sum_{i\in{\text{out}\atop\text{in}}}  \o_i  \frac{ [ -  {\hat p}_i \cdot {\hat q}(x) ]^{1+i\nu} }{ i\nu ( 1+i\nu )  }  \;  , \\
\CJ_{ab}(x) &= \lim_{\nu \to 0} [ \CJ^+_{ab}(\nu,x) + \CJ^-_{ab}(\nu,x)] \; . 
\end{split}
\end{equation}

\section{Magnetic Charges}\label{sec:MagSoft}

The previous section reviewed soft theorems and infrared divergences for electrically charged states in gauge theories. In this section, we generalize those results to include magnetically charged or dyonic states. The first discussion of an abelian magnetic soft theorem appeared in \cite{Strominger:2015bla}. The results on the chromo-magnetic soft theorem and the soft theorem for extended objects are new.

\subsection{Abelian Gauge Theories}
\label{Sec:MagSoft}

In four dimensions, zero-dimensional particles can carry both electric and magnetic charges. The magnetic charge is defined by the surface integral
\begin{equation}
\begin{split}
P = \frac{1}{2\pi} \int_{S^2} F \;  
\end{split}
\end{equation}
taken at spatial infinity. The generalized soft theorems and infrared divergences for magnetic states can be obtained using the electromagnetic duality transformation 
\begin{equation}
\begin{split}\label{EMduality}
F \to {\wt F} = - \frac{2\pi}{e^2} \star F \; , \qquad e \to {\wt e} = \frac{2\pi}{e}\;  , \qquad Q \to {\wt Q} = P  \; , \qquad P \to {\wt P} = - Q\; . 
\end{split}
\end{equation}
The allowed set of magnetic charges is constrained by the Dirac quantization condition $Q_i P_j  \in \mzz$. This implies that magnetic charge is quantized in multiples of $m_0\in\mzz$,  the smallest magnetic charge in the spectrum:
\begin{equation}
P_i = m_0 w_i \; , \qquad w_i \in \mzz \; .
\end{equation}
The integer $m_0$ is theory-dependent.

To write down the soft theorem for magnetically charged particles, we Fourier transform the first equation of \eqref{EMduality} to determine the dual polarization vector:
\begin{equation}
\begin{split}
q_{[\mu} {\wt \ve}_{\nu]} (q)  = - \frac{\pi}{e^2} \e_{\mu\nu\rho\s} q^{[\rho} \ve^{\s]} (q) \quad \implies \quad {\wt \ve}_\mu(q) = \frac{2\pi}{e^2} \e_{\mu\nu\rho\s}  \frac{n^\nu q^\rho }{n\cdot q} \ve^\s(q) \; . 
\end{split}
\end{equation}
In the basis \eqref{gluonpol}, this explicitly evaluates to
\begin{equation}
\begin{split}
{\wt \ve}_a(x) = \frac{2\pi}{e^2} \e_{ab} \ve^b(x) \; , 
\end{split}
\end{equation}
where $\epsilon_{ab}$ is the 2D Levi-Civita tensor normalized as $\epsilon_{12} = \epsilon^{12} = 1$. Magnetically charged particles couple to the dual polarization of the photon so the soft photon theorem \eqref{soft-photon-thm} becomes \cite{Strominger:2015bla}
\begin{equation}
\begin{split}\label{soft-photon-thm-mag}
\avg{ \CO_a(\o,x) \CO_1 \cdots \CO_n } ~~ \stackrel{\o\to0}{\longrightarrow} ~&~ e \sum_i {\wt Q}_i \frac{ {\hat p}_i \cdot {\wt \ve}_a (x) }{ {\hat p}_i \cdot {\hat q}(x)  } \avg{  \CO_1 \cdots \CO_n }  \\
&= \frac{2\pi}{e} \e_{ab} \sum_i P_i \frac{ {\hat p}_i \cdot \ve^b (x) }{ {\hat p}_i \cdot {\hat q}(x)  }  \avg{  \CO_1 \cdots \CO_n }\;   . 
\end{split}
\end{equation}
In the language of conformal correlators, this reads
\begin{equation}
\begin{split}\label{Sains-mag}
\avg{ S_a(x) \CO_1 \cdots \CO_n } = {\wt \CJ}_a (x) \avg{ \CO_1 \cdots \CO_n  }\; ,  \qquad  {\wt \CJ}_a (x) \equiv \frac{2\pi}{e^2} \epsilon_{ab} \p^b \sum_i P_i \log [ - {\hat p}_i \cdot {\hat q}(x) ] \; .
\end{split}
\end{equation}

This analysis extends to dyonic states in a straightforward manner. For any two pairs of dyons with charges $(Q_i,P_i)$ and $(Q_j,P_j)$, the Dirac-Schwinger-Zwanziger quantization condition (derived by quantizing the total angular momentum of the electromagnetic field of a two-dyon state \cite{Goddard_1978}) implies
\begin{equation}
\label{SZquantcond}
Q_i P_j - Q_j P_i \in \mzz \; .
\end{equation}
It follows that the charge spectrum for dyons takes the form \cite{Olive:1995sw}\footnote{The generalized $SL(2,\mzz)$ transformation acts as $\tau = \tau_1 + i \tau_2  \to \frac{ a \tau + b }{ c \tau + d }$, $n_i \to a n_i - b w_i$, $w_i \to - c n_i + d w_i$ with $a,b,c,d\in\mzz$ and $ad-bc=1$. On the field strength this acts as $F + i \star F \to ( c \tau + d ) ( F + i \star F )$.}
\begin{equation}
\begin{split}\label{chargesol}
Q_i = n_i + \frac{\vt}{2\pi} w_i \; , \qquad P_i = m_0 w_i \; , \qquad n_i,w_i\in\mzz \; , \qquad \vt \in [ 0 , 2\pi )\; .
\end{split}
\end{equation}
The soft theorem for these dyons simply combines \eqref{Sains} and \eqref{Sains-mag} in a duality covariant way
\begin{equation}
\begin{split}\label{Sains-dyon}
\avg{ S_a(x) \CO_1 \cdots \CO_n } &=  [ \CJ_a(x) + {\wt \CJ}_a(x) ] \avg{ \CO_1 \cdots \CO_n  } \\
&= \sum_i  ( n_i \d_{ab} +  w_i  \tau_{ab} )  \p^b \log [ - {\hat p}_i \cdot {\hat q}(x) ] \avg{ \CO_1 \cdots \CO_n  } \; ,
\end{split}
\end{equation}
where
\begin{equation}
\label{taudef}
\tau_{ab} \equiv \tau_1 \d_{ab}  + \tau_2 \e_{ab} \; , \qquad \tau_1 \equiv \frac{\vt}{2\pi}   \; , \qquad \tau_2 \equiv \frac{2\pi m_0}{e^2}   \; . 
\end{equation}
In addition to the electric current defined in \eqref{eq:CurrentShadow}, we can now also define a magnetic current which couples to magnetic particles,
\begin{equation}
\begin{split}\label{Kacurrdef}
K_a \equiv  -(\star J)_a =-\e_{ab} J^b \; . 
\end{split}
\end{equation}
This satisfies
\begin{equation}
\begin{split}
\avg{ \p^a K_a (x) \CO_1 \cdots \CO_n } &= \frac{2\pi}{e^2} \sum_i P_i \, \CK_2 ( m_i / \o_i , x_i ; x )  \avg{\CO_1 \cdots \CO_n  } \; .
\end{split}
\end{equation}

The structure of infrared divergences in the presence of magnetic charges is similarly obtained by replacing $\CJ_a \to \CJ_a + {\wt \CJ}_a$ in the integral \eqref{Adefabelian}. The infrared factor is $\G_\text{ph} = \a ( A_1 + 2\pi i A_2 )$ with
\begin{equation}
\begin{split} \label{eq:MagDiv}
A_1 &=   \int  \frac{d^d x}{(2\pi)^d} \left[ \CJ_a(x) + {\wt \CJ}_a(x) \right]^2 \;  , \\
A_2 &= \int_{-\infty}^{\infty} \frac{d\nu}{2\pi}  \int  \frac{d^d x}{(2\pi)^d} \bigg( | \CJ_a^+(\nu,x) + {\wt \CJ}^{+}_a (\nu,x) |^2   +  | \CJ_a^-(\nu,x) + {\wt \CJ}^{-}_a (\nu,x) |^2  \bigg)_r \; , 
\end{split}
\end{equation}
where
\begin{equation}
\begin{split}
{\wt \CJ}_a^{\pm} (\nu,x) \equiv \frac{2\pi}{e^2} \e_{ab} \p^b \sum_{i\in{\text{out}\atop\text{in}} } P_i  \frac{ [ -  {\hat p}_i \cdot {\hat q}(x) ]^{i\nu}}{i\nu} \; . 
\end{split}
\end{equation}
This formula is straightforward for purely magnetic scattering and follows from an application of the duality transformation \eqref{EMduality}. The case with both electric and magnetic charges is slightly more subtle. The quantization condition \eqref{SZquantcond} implies that there is no small expansion parameter when an electron scatters off of a monopole: the effective coupling is always of order one. Weinberg also noted \cite{Weinberg:1965rz} that the single photon exchange diagram between an electron and a monopole appears to violate Lorentz invariance. Both of these complications are resolved by explicitly exponentiating the diagrams in figure \ref{IRdiagrams}. Although the coupling is of order one, it is possible to treat the infrared exchange to all orders to obtain a reliable approximation. Similarly, since all photon exchange diagrams contribute at the same order in the $e^2$ expansion, it is only their sum which must be (and is \cite{Terning:2018udc}, for appropriately quantized charges) Lorentz and gauge invariant.

\subsection{Non-abelian Gauge Theories and Chromodyons}\label{MagSoftGluon}
The analog of the Dirac quantization condition in non-abelian gauge theory is the GNO constraint \cite{Goddard:1976qe}. For a given ``electric gauge group'' $G$, this condition restricts the allowed magnetic charges of the theory to be weights for the dual group $G^\vee$ obtained from $G$ by taking the dual root lattice.

In a deconfined non-abelian gauge theory, there is no difficulty in identifying the electric gauge group $G$ in the spectrum of electrically charged states. Similarly, the magnetic gauge group $G^\vee$ has an unambiguous action in the sector of purely magnetically charged states. The essential complication arises in understanding the action of $G$ and $G^\vee$ in the dyonic sector of the theory. Indeed, semiclassical quantization of a single isolated non-abelian monopole does not yield states transforming in representations of $G$ \cite{Abouelsaood:1982dz,Nelson:1983em} as one would expect if the asymptotic symmetry group was truly $G\times G^\vee$. Chromo-magnetic charge apparently spoils the action of $G$ in the dyonic sector of the Hilbert space and the global part of the electric gauge group is ill-defined \cite{ Balachandran:1982gt, Nelson:1983bu, Balachandran:1983xz, Balachandran:1983fg, Nelson:1983fn}.

There is a simple physical explanation of this mathematical fact. The long-range chromo-magnetic field of a monopole is controlled by a Gauss law. In the magnetically charged sector of the theory, the magnetic charge 
\begin{equation}
Q^I=\frac{1}{2\pi}\int_{i^0} F^I
\end{equation}
is non-vanishing which implies that parallel transport on the sphere at spatial infinity is path dependent. The ``global part of the electric gauge group'' is defined by performing the \emph{same} large gauge transformation everywhere on the celestial sphere. 
To compare the gauge parameter at two separate points on the sphere, we are required to parallel transport from one point to the other. In the presence of asymptotic chromo-magnetic flux the parallel transport is path dependent (for directions in color space that do not commute with the asymptotic value of $F$) and so the definition of the global gauge group is ambiguous.\footnote{This is essentially the gauge theory analogue of the ``problem of angular momentum'' in general relativity and one might hope that it is resolved analogously.}  However, if there is no net magnetic charge then the electric group $G$ can be unambiguously defined. Indeed, semiclassical quantization of a monopole-antimonopole pair produces states that do form representations of the electric color group \cite{Nelson:1983fn}. Therefore, provided that the long-range magnetic field vanishes in the \emph{in} and \emph{out} states (i.e. both have net zero magnetic charge), it is possible to consider a $S$-matrix involving magnetically charged states (the non-abelian magnetic analog of $e^+e^-$ scattering). It is in this restricted case that we expect the following chromo-dyon soft theorem to apply.

We will explore the magnetic soft gluon theorem guided by analogy with the abelian case\footnote{Since the equations of motion in non-abelian gauge theory explicitly involve the electric gauge potential, the simple duality transformation \eqref{EMduality} must be more complicated but might hold asymptotically or in the soft limit. For this reason we consider the formulas in this subsection conjectural.   } and the electromagnetic duality transformation \eqref{EMduality}. We conjecture the following form of the magnetic soft gluon theorem
\begin{equation} 
\avg{ S^I_a(x)   \CO_1 \cdots \CO_n } = \frac{2\pi i}{g_{\ym}^2}  \e_{ab} \p^b \sum_i  \log [ - {\hat p}_i \cdot {\hat q}(x) ]  {\hat T}_i^I \avg{   \CO_1 \cdots \CO_n } \; .
\end{equation}
Here ${\hat T}^I$ is a Lie algebra generator of $G^\vee$ in representation $R_i$ under which the particle transforms. Generalizing to chromodyons and including the $\vt$-term contribution, the generalized soft gluon theorem can be written as
\begin{equation}
\begin{split}
\avg{ S^I_a(x) \CO_1 \cdots \CO_n } &=  i \sum_i ( T_i^I \d_{ab}  +  {\hat T}^I_i \tau_{ab} ) \p^b \log [ - {\hat p}_i \cdot {\hat q}(x) ] \avg{ \CO_1 \cdots \CO_n  }  \; .
\end{split}
\end{equation}
We are unaware of any results regarding the scattering of electric and magnetic states in non-abelian gauge theory beyond the non-relativistic approximation of motion on moduli space. It would be interesting to understand the IR divergence structure of such amplitudes.

\subsection{Soft Theorem for Magnetic Branes in $d>2$}
\label{Sec:MagSoftd}

In four-dimensional gauge theory, both the vector potential $F\equiv dA$ and the dual vector potential $\star F \equiv d\tilde{A}$ are one-forms and they couple to zero-dimensional excitations with one-dimensional world-lines. In higher dimensions the vector potential remains a one-form and couples to particles, while the dual potential couples to extended objects with $(d-1)$-dimensional world-volumes which we will refer to as magnetic branes. In $\mathbb{R}^{1,d+1}$, the dual magnetic potential is a $(d-1)$-form defined by
\begin{equation}
\begin{split}
{\tilde F}_{(d)} = d {\tilde A}_{(d-1)} =  \star F_\2 = \star d A_\1 \; . 
\end{split}
\end{equation}
Here, we have introduced a subscript to make the rank of the differential form explicit. The magnetic brane couples to the gauge field through the Bianchi equation
\begin{equation}
\begin{split}\label{magBianchihigherd}
d F_\2 = 2\pi \star {\tilde J}_{(d-1)} 
\end{split}
\end{equation}
where ${\tilde J}_{(d-1)}$ is the conserved $(d-1)$-form current. Classically, this current is given by\footnote{See for example eq. (18.40) in \cite{Ortin:2015hya}.}
\begin{equation}
\begin{split}
\label{eq:ExtendedCurrent}
{\tilde J}_{(d-1)}^{\mu_1\dots \mu_{d-1}}(x) = \sum_i P_i \int_{\Sigma_i} dX_{\S_i}^{\mu_1} \w \cdots \w dX_{\S_i}^{\mu_{d-1}}\delta^{(d+2)}( x , X_{\S_i}) \;  , \qquad d \star {\tilde J}_{(d-1)} = 0 \; . 
\end{split}
\end{equation}
The integral is over the world-volume $\S_i$ of the branes and $X_{\S_i} : \S_i \hookrightarrow \CM$ describes the embedding of the world volume $\S_i$ in spacetime. $P_i$ is the magnetic charge of the branes.

The soft theorem in the presence of such branes can be derived from asymptotic symmetries \cite{Strominger:2017zoo}. In any dimension, the usual soft photon theorem \eqref{soft-photon-thm} is related to the conservation of a $(d+1)$-form
\begin{equation}
\begin{split}
\star J_\ve \equiv \frac{1}{e^2} d ( \ve_\0 \star F_\2 ) \; , \qquad d \star J_\ve = 0 \; . 
\end{split}
\end{equation}
Using Maxwell's equation, we can write the current as
\begin{equation}
\begin{split}
\star J_\ve \equiv \frac{1}{e^2}  d \ve_\0 \w \star F_\2 +  \ve_\0 \star J_\1 \; , 
\end{split}
\end{equation}
where $J_\1$ is the usual electric current of the model. The charge associated to this current generates electric large gauge transformations under which $A_\1 \to A_\1 + d \ve_\0$. Evaluating this charge on the \emph{in}-state and the \emph{out}-state, one can derive the leading soft photon theorem \cite{He:2014cra,He:2019jjk}. The first term above inserts a soft photon with polarization $d\ve_\0$ in the $S$-matrix whereas the second term generates the soft factor in \eqref{soft-photon-thm} via a Fourier transform
\begin{equation}
\begin{split}
  i e \ve_\mu (q)\int d^{d+2} x e^{- i q \cdot x } J_\1^\mu ( x )  ~ \stackrel{q \to 0}{\longrightarrow} ~ e \sum_i Q_i \frac{ p_i \cdot \ve(q) }{ p_i \cdot q  } \; .
\end{split}
\end{equation}
To derive the magnetic soft theorem, we define the magnetic conserved current
\begin{equation}
\begin{split}
\label{eq:HigherMagCurrent}
\star {\tilde J}_{\tilde \ve} \equiv  \frac{1}{2\pi} d( {\tilde \ve}_{(d-2)} \w F_\2 ) \; , \qquad d \star {\tilde J}_{\tilde \ve} = 0 \; . 
\end{split}
\end{equation}
Using the Bianchi equation \eqref{magBianchihigherd}, this can be written as
\begin{equation}
\begin{split}\label{astJdef}
\star {\tilde J}_{\tilde \ve} = \frac{1}{2\pi} d  {\tilde \ve}_{(d-2)} \w F_\2  + (-1)^d  {\tilde \ve}_{(d-2)} \w \star {\tilde J}_{(d-1)} \; .
\end{split}
\end{equation}
The charge associated to this current generates magnetic large gauge transformations under which ${\tilde A}_{(d-1)} \to {\tilde A}_{(d-1)}  + d {\tilde \ve}_{(d-2)}$ \cite{Aurilia:1993qi,Afshar:2018apx,He:2019ywq}. The corresponding Ward identity for this symmetry is obtained by inserting the associated  charge into the $S$-matrix. The first term in \eqref{astJdef} inserts a dual soft photon with the $(d-1)$ form polarization $d{\tilde \ve}_{(d-2)}$. This polarization is related to the polarization vector as
\begin{equation}
\label{eq:MagPolHigher}
( {\tilde \ve}_a ) _{\mu_1\dots \mu_{d-1}}(q) = (-1)^d \frac{2\pi}{e^2} \e_{\mu_1\dots \mu_{d-1}\nu \rho \s} \frac{n^\nu q^\rho }{n\cdot q} \ve_a^\s(q)\;  .
\end{equation}
The second term generates the soft factor via a Fourier transform
\begin{equation}
\begin{split}
&i e ({\tilde \ve}_a)_{\mu_1\dots \mu_{d-1}}(q)  \int d^{d+2} x e^{- i q \cdot x } {\tilde J}_{(d-1)}^{\mu_1 \cdots \mu_{d-1} }  ( x )  ~ \stackrel{q \to 0}{\longrightarrow} ~  \frac{2\pi}{e} \sum_i P_i \,  S^{\text{brane}}_{i,a}(q) \; ,
\end{split}
\end{equation}
where
\begin{equation}
\begin{split}\label{Sidef}
 S^{\text{brane}}_{i,a}(q) =  i  (-1)^d (d-1)! \frac{ \ve_{a\mu}(q) n_\nu q_\rho }{n\cdot q}  \int_{\Sigma_i}  \star ( d X_{\S_i}^\mu  \w d X_{\S_i}^\nu \w d X_{\S_i}^\rho )  \, e^{- i q \cdot X_{\S_i} } \;  .
\end{split}
\end{equation}
We evaluate this quantity in a special case. At late times, we can assume that the magnetic branes are non-interacting and are freely moving. We can describe a brane in terms of a timelike parameter $\tau$ and $d-2$ spatial parameters $\s^\a$, $\a=1,\cdots,d-2$ as
\begin{equation}
\begin{split}
X_{\S_i}^\mu = \chi_i^\mu + \tau p_i^\mu + r^\mu_i(\s) \; . 
\end{split}
\end{equation}
$\chi_i$ describes the center of mass of the brane and $p_i^\mu$ describes the center of mass momentum. The soft factor for such a brane takes the form
\begin{equation}
\begin{split}\label{branesf}
S^{\text{brane}}_{i,a}(x) = - \frac{1}{\o} \frac{2(d-1)! }{ {\hat p}_i \cdot {\hat q}(x) }  {\hat p}_i^\mu n^\nu {\hat q}^\rho \p_a {\hat q}^\s  \e_{\mu\nu\rho\s\mu_1 \cdots \mu_{d-2} } \int d^{d-2} \s    e^{ - i q \cdot r_i (\s)  } \p_{\s^1} r_i^{\mu_1}  \cdots  \p_{\s^{d-2} } r_i^{\mu_{d-2} } \;  .
\end{split}
\end{equation}
This suggests that the soft factor has a simple pole at $\o  = 0$. However, if the brane has a non-compact spatial volume then the integral can produce additional factors of $\o^{-1}$. To see this, consider a brane localized along the $(x^3,\cdots,x^d)$ directions. We describe this by the spatial vector $r^\mu(\s) = (0,0,0,\s^1,\cdots,\s^{d-2},0)$. In this case, the soft factor works out to be
\begin{equation}
\begin{split}\label{branesfspecial}
S^{\text{brane}}_{a_\perp} (x) =  \frac{(d-1)!}{\o^{d-1} } \e_{a_\perp b_\perp} \p^{b_\perp} \log [ - {\hat p} \cdot {\hat q}(x_\perp) ]   \d^{(d-2)} ( x_\| ) \;  , \qquad S^{\text{brane}}_{a_\|} (x) = 0 \; , 
\end{split}
\end{equation}
where $x^a_\| = (0,0,x^3 ,\cdots,x^d)$ and $x^a_\perp = ( x^1 , x^2 , 0 , \cdots , 0)$ describe the longitudinal and transverse directions to the brane and $a_\perp=(1,2)$, $a_\| = (3,\cdots,d)$ are the corresponding indices. 

The soft factor therefore diverges as $\o^{-(d-1)}$. This leading divergence can be extracted by defining a new soft mode
\begin{equation}
\begin{split}
S^\dual_a(x) \equiv \frac{1}{e} \oint_\CC \frac{d\o}{2\pi i} \o^{d-2} \CO_a(\o,x) \;  . 
\end{split}
\end{equation}
This is the $n=d-1$ mode in the class of Mellin transformed operators defined by the compact contour \eqref{compactcontour} and has scaling dimension $\D = d-1$. The magnetic soft theorem in $d>2$ reads
\begin{equation}
\begin{split}\label{magsoftthmhigherd}
\avg{ S^\dual_a(x) \CO_1 \cdots \CO_n } = \frac{2\pi}{e^2} \sum_{i\,\in\,\text{brane}} P_i \, S^{(0)\text{brane}}_{i,a}(x)  \, \avg{ \CO_1 \cdots \CO_n } \; , 
\end{split}
\end{equation}
with
\begin{equation}
\begin{split}
S^{(0)\text{brane}}_{i,a}(x)  = \lim_{\o \to 0} \o^{d-1} S^{\text{brane}}_{i,a}(x) \; . 
\end{split}
\end{equation}
It should be possible to construct a higher-form current (sourced by extended objects in CFT$_d$) from this soft operator following the construction of \eqref{Jains}. We leave this for future work.

\section{A Boundary Model for Bulk Infrared Physics}\label{sec:Actions}

In this section, we exploit the asymptotic symmetry structure of gauge and gravitational theories to construct a boundary ($d$-dimensional) theory whose correlators reproduce bulk ($(d+2)$-dimensional) infrared physics. Scattering amplitudes in the bulk theory are calculated by a path integral 
\begin{equation}
\begin{split}
\avg{ \CO_1 \cdots \CO_n }_\mu = \int [d \varphi ] e^{i S_{\text{bulk}} [ \varphi ] } \CO_1 \cdots \CO_n \; ,
\end{split}
\end{equation}
where $\varphi$ collectively denotes all the fields in the theory (including any massless photons, gluons or gravitons). The subscript $\mu$ denotes the IR cut-off used to regulate and evaluate the path integral on the RHS (all fields in the path integral are taken to have $\o > \mu$).

In order to isolate the infrared physics in the path integral, we separate all fields into a soft piece $\varphi_s$ (which has $\mu < |\o| < \L$) and a hard piece $\varphi_h$ (which has $|\o| > \L$). The scales $\mu$, $\L$ are taken to be much smaller than the gap in the massive sector so the soft fields $\varphi_s$ only consist of low energy massless fields. In models with hard-soft factorization we can integrate out the hard fields
\begin{equation}
\begin{split}\label{hardpathintegral}
\avg{ \CO_1 \cdots \CO_n }_\mu &= \int [d \varphi_{s}  ] \int [ d \varphi_{h} ] e^{ i S_{\text{bulk}} [ \varphi_s , \varphi_h ] } \CO_1 \cdots \CO_n \\
&= \avg{ \CO_1 \cdots \CO_n }_\L \int  [d \varphi_{s} ] e^{ - S_{\text{soft}} [ \varphi_s ] - S_{\text{int}} [ \varphi_s , j ]  }  \;  .
\end{split}
\end{equation}
The integral over the hard modes produces the hard amplitude $\avg{ \CO_1 \cdots \CO_n }_\L$ which does not depend on any of the soft fields or the infrared cutoff $\mu$. The left-over integral over the soft modes contains an effective action which we have separated into two parts -- $S_{\text{soft}}$ describes the self-interactions of the soft modes and $S_{\text{int}}$  describes the coupling of the long-wavelength fluctuations to the hard external particles in the scattering amplitude. This interaction is encoded through a classical (non-fluctuating) hard matter current $j$ whose structure is independent of the microscopic details of the  theory. Comparing to the general structure of infrared divergences \eqref{IRgen}, one finds
\begin{equation}
\begin{split}\label{IR-res-soft}
 \int  [d \varphi_{s} ] e^{ - S_{\text{soft}} [ \varphi_s ] - S_{\text{int}} [ \varphi_s , j ]  } = e^{ - \G}  \; .
\end{split}
\end{equation}
The above analysis assumes that the operators $\CO_i$ are all chosen from the hard sector. Scattering amplitudes with external soft particles require additional soft insertions $S_i(x)$ in the path integral. The multi-soft theorems guarantee that the corresponding amplitude factorizes so we require
\begin{equation}
\begin{split}\label{IR-res-soft-1}
 \int  [d \varphi_{s} ] e^{ - S_{\text{soft}} [ \varphi_s ] - S_{\text{int}} [ \varphi_s , j ]  } S_1 \cdots S_m = e^{ - \G}  \CJ_1 \cdots \CJ_m  \; ,
\end{split}
\end{equation}
where $S_i$ denotes a generic soft insertion and $\CJ_i$ denotes the corresponding soft factor.

As we will show in the rest of this section, the action for the soft modes and their coupling to the matter sources is essentially fixed by symmetry considerations. The coupling constants of the path integral on the LHS of \eqref{IR-res-soft-1} are completely fixed by the soft theorem and the one-loop infrared divergence. It appears nontrivial that the intrinsically Euclidean path integral \eqref{IR-res-soft-1} is capable of reproducing intrinsically Lorentzian infrared dynamics.

\subsection{Broken Symmetries in Celestial CFT}

Symmetries in quantum field theory can be approximate or exact. If they are exact, they can be spontaneously broken or unbroken, depending on the nature of the vacuum state. In some cases, it is useful to treat an approximate symmetry as if it were exact. If the model with the exact symmetry lies in the broken phase and if the source of explicit symmetry breaking is small, then the pseudo-Goldstone bosons (along with any other massless fields) can be used to reliably approximate the low-energy dynamics.

There is by now a standard set of claims in the literature (see \cite{Strominger:2017zoo} and references therein) regarding the symmetry structure of gauge theory and quantum gravity in asymptotically flat space. Gauge transformations or diffeomorphisms with non-compact support that respect appropriate boundary conditions act non-trivially on the Hilbert space in these models. We denote this infinite-dimensional asymptotic symmetry group of the theory by $\CG$. Gauge transformations or diffeomorphisms which are asymptotically ``constant'' (in an appropriate sense) represent the global part of $\CG$ which we denote by $G$. The existence of the $\CG$ symmetry implies an infinite degeneracy of vacuum states in the model. These vacuum states are preserved by $G$ but not by $\CG$. It follows that in these models, the exact symmetry $\CG$ is spontaneously broken down to $G$. The low energy dynamics of these models are described by the corresponding Goldstone bosons.
 
It is important to remember that the group $\CG$ and its subgroup $G$ are only defined in spacetimes with non-compact Cauchy slices. There are no charged states for $G$ or Goldstone bosons for $\CG/G$ on a compact manifold, just as there is no spontaneous symmetry breaking of ordinary symmetries in finite volume. In order to recover $\CG$ and $G$ from an infinite volume limit, it is important to take a limit of space-like slices \textit{with} boundary. In this limit, the corresponding edge modes become the Goldstone modes for $\CG/G$. This point is important precisely because gauge theories suffer from infrared issues. Perturbative calculations in gauge theory and gravity require the regulation of long-wavelength fluctuations. Any method of regulating the infrared divergences (e.g. a small photon mass, $d>4$ dimensional regularization, finite volume regularization, etc.) explicitly breaks the $\CG$ symmetry.\footnote{In the case of dimensional regularization, the group itself changes.}  The effective action for these Goldstone modes in the regulated theory will therefore contain symmetry breaking terms, and we are forced to consider pseudo-Goldstone bosons associated to a symmetry with both spontaneous and explicit breaking.\footnote{This is also the case in JT gravity, which describes the pseudo-Goldstone bosons associated to the IR divergence in AdS$_2$ \cite{Maldacena:2016upp}.} If the symmetry breaking terms vanish as the regulator is removed, then we expect to land on a model with an exact $G$ symmetry and Goldstone modes for the coset  $\CG/G$. This will be the case when $d>2$ and when the gauge theory is in the Coulomb phase. However, as we will see, when $d=2$ the pseudo-Goldstone modes become strongly coupled as the regulator is removed and we must keep the leading symmetry breaking terms in the effective description. This strong coupling behavior is a two-dimensional reflection of the infrared divergence in $D=4$ spacetime dimensions. The four-dimensional Goldstone bosons are strongly coupled because two-dimensional Goldstone bosons are strongly interacting.

Since the celestial CFT$_d$ formalism is a repackaging of the bulk $(d+2)$-dimensional dynamics, we expect the asymptotic symmetry structures to constrain the $d$-dimensional model. When the bulk theories are in the Coulomb phase, we expect  a global $G$ symmetry in the CFT$_d$ and a corresponding conserved current. These are precisely the operators \eqref{eq:CurrentShadow}, \eqref{eq:YMCurrentShadow} and \eqref{eq:TranslationCurrent} constructed from the shadow transforms of the soft operators. Their existence is required in order to match global symmetries and the universal nature of soft limits guarantees that they generate the correct symmetry transformations. We also expect to be able to phrase the dynamics of the $\CG/G$ Goldstone bosons simply in $d$-dimensional language since it is also controlled by symmetries.

\subsection{$U(1)$ Gauge Theory}

We will begin by applying the above discussion to the case of $U(1)$ gauge theories. Goldstone actions for infinite-dimensional symmetry groups like $\mathcal{G}$ are somewhat exotic, so we begin by reviewing the finite-dimensional case of global $U(1)$ symmetry breaking and then proceed by analogy.

The simplest model of a spontaneously broken global $U(1)$ symmetry involves a scalar field with a Lagrangian invariant under the transformation $\phi(x) \to e^{ i Q \ve } \phi(x)$. If the scalar potential has a global minimum with $|\phi_{\text{min}}|=v\neq 0$, then the boundary value of the field at spatial infinity cannot vanish and the field acquires a vacuum expectation value set by the boundary condition at infinity: $\avg{ \phi(x) } = v \neq 0$. In this case $\phi(x)$ is the order parameter, and its boundary condition spontaneously breaks the $U(1)$ symmetry of the model.
The normalizable fluctuations about this vacuum are described by the field redefinition $\phi(x) = ( v + h(x) ) e^{ i Q \t (x) / v }$. Excitations of $h(x)$ are gapped while the Goldstone mode $\t(x)$ is massless and protected by the non-linearly realized $U(1)$ symmetry $\t(x) \to \t(x) + \ve$. At low energies, only the Goldstone mode survives and the dynamics is described by an action of the form $ \int  [ - \frac{1}{2} ( \p \t )^2 + \cdots ]$ which is constrained to be invariant under the $U(1)$ symmetry. If the symmetry is also (weakly) explicitly broken in the ultraviolet, then one must add small symmetry breaking terms to this action, e.g. a mass term for $\t$.

In the case of a $U(1)$ gauge theory, the relevant symmetry group is the ``asymptotic symmetry group'' $\mathcal{G}$, also known as the group of ``large'' gauge transformations with non-compact support \cite{He:2014cra, Campiglia:2015qka, Kapec:2014zla, Kapec:2015ena, Campiglia:2018dyi}. Elements of $\mathcal{G}$ correspond to maps $S^{d}\to U(1)$ and the relevant order parameter is a Wilson line on the celestial sphere
\begin{equation}
\begin{split}
W(x) \equiv \exp \left( i \int_{x_0}^{x} C \right) \; , \qquad C = A |_{\p \ci} \; . 
\end{split}
\end{equation}
Here $x_0$ is an arbitrarily chosen base point. This order parameter is invariant under constant gauge transformations, but rotates with a phase under generic elements of $\mathcal{G}$:
\begin{equation}
\begin{split}
W(x) ~ \stackrel{\CG}{\longrightarrow}~ e^{ i ( \ve(x) - \ve(x_0) )} W(x) \;  , \qquad \ve(x) \sim \ve(x) + 2\pi \; .
\end{split}
\end{equation}
When the quantum theory is in the Coulomb phase, this order parameter is non-vanishing, $\avg{ W(x) } \neq 0$, which indicates that the $\CG$ symmetry is spontaneously broken down to $G$. The Goldstone mode for the broken symmetry is $C$ and it transforms non-linearly as $C \to C + d \ve$.

We now turn to the structure of the low energy action, $S_{\text{soft}}$, which describes the self-interactions of the soft modes. The first thing to note is that the soft photon operator $S$ in \eqref{Sdef} is independent of the Goldstone mode $C$. In particular, while $C$ transforms under large gauge transformations, $S$ is related to gauge-invariant matrix elements and therefore does not transform. Since the operator $S$ creates a soft photon, it survives in the low energy limit so $S_{\text{soft}}$ depends on both $C$ and $S$. We will begin by considering the case in which all external states in the scattering amplitude are hard. In this case, we can integrate out $S$ and obtain an effective action for the mode $C$. Having determined this effective action, we will then ``integrate in'' the mode $S$ by consideration of the soft photon theorem. 

In the absence of magnetic charges the gauge field edge mode  $C$ is flat and realizes the $\mathcal{G}$ symmetry $C \to C + d\varepsilon$ non-linearly.  When the $\mathcal{G}$ symmetry is exact but spontaneously broken, the effective action for $C$ must be invariant under this transformation.  What sort of action is invariant under the infinite-dimensional symmetry group $\mathcal{G}$? The simplest example is of course a model with $d$-dimensional abelian gauge invariance in which we \textit{do not quotient by gauge transformations}. 
Since $C$ is flat, we do not want to integrate over all abelian connections, only those which are flat. A $d$-dimensional $BF$ theory has the appropriate domain of integration and the correct symmetries\footnote{This $BF$ theory differs from the models usually considered in both form and function. It does not involve the usual quotient by ``small gauge transformations,'' but is viewed as part of a higher-dimensional theory.}
\begin{equation}
\begin{split}
\label{eq:BF}
S_{BF}[B,C] =i \int  B\wedge F \; , \qquad F = d C \; . 
\end{split}
\end{equation}
Integrating out $B$ sets $F=0$ and the soft dynamics are effectively trivial. However, as we have noted previously, the infrared regulator explicitly breaks the $\CG$ symmetry so the effective action can also contain symmetry breaking terms. These correspond to non-gauge invariant interactions that depend explicitly on $C$ rather than $F$. 

In standard constructions of effective actions for pseudo-Goldstone bosons, locality is an organizing principle. In the case at hand, we are attempting to construct an exotic ``interdimensional effective field theory'' that calculates bulk  $(d+2)$-dimensional quantities in terms of $d$-dimensional field variables. While the bulk theory must be local, the boundary theory need not be. Indeed, given that the boundary theory is required to reproduce long wavelength effects in the bulk, the theory could indeed be non-local. With this in mind, the most general leading symmetry breaking term that we can add to the action takes the form
\begin{equation}
\begin{split}\label{eq:AbelianBreaking}
S_{\text{soft}}[C] = \int d^d x d^d y \, (P^{-1})^{ab}(x-y) C_a(x) C_b(y) \; .
\end{split}
\end{equation}
$P_{ab}(x-y)$ is the propagator for $C$ and will be determined in the next section. The choice $P_{ab}=\delta_{ab}$ would be a mass term for the $d$-dimensional gauge field.  Integrating out the $B$ field still enforces the flatness constraint
\begin{equation}
\begin{split}\label{C-flat-theta}
C = d \t \; , \qquad \t(x) \sim \t(x) + c \; , 
\end{split}
\end{equation}
and the mass term for the gauge field becomes a kinetic term for the gauge parameter $\theta(x)$. The identification is required because the global $G$ symmetry is unbroken on the vacuum: $C$ is a Goldstone with target $\mathcal{G}/G$ so the zero mode of $\theta(x)$ is a redundant parameterization of the space of flat connections and should not be integrated over. Under large gauge transformations,
\begin{equation}
\begin{split}\label{theta-LGT}
\t(x) ~~ \stackrel{\CG}{\longrightarrow} ~~  \t(x) + \ve(x) \;.
\end{split}
\end{equation}
The action \eqref{eq:AbelianBreaking} explicitly breaks the $\mathcal{G}$ symmetry but is invariant under the global $U(1)$ transformations.

Next, we turn to the form of the interaction term $S_{\text{int}}$. The factorized structure of  scattering amplitudes suggests that the scattering operators $\CO_i$  factorize into a soft and a hard piece:
\begin{equation}
\begin{split}
\CO_i ( \o_i , x_i ) = U_i[\t] \CO_i^H(\o_i,x_i) \; . 
\end{split}
\end{equation}
The hard operators $\CO^H$ depend only on the hard fields $\varphi_h$ and decouple from the soft path integral \eqref{hardpathintegral}, although they obviously contribute to the hard scattering amplitude. The soft operator $U_i[C]$ can be determined by considering the large gauge transformation of $\CO_i$ which takes the form\footnote{We assume for now that $\CO_i$ is purely electrically charged. The generalization to incorporate dyons will be considered in section \ref{sec:magcharge2d}.}
\begin{equation}
\begin{split}\label{Gtransform-U}
\CO_i ( \o_i , x_i ) ~~ \stackrel{\CG}{\longrightarrow}~~ \exp \left[ i Q_i \int d^d x \ve(x) \CK_d ( m_i / \o_i , x_i ; x ) \right] \CO_i ( \o_i , x_i ) \; .
\end{split}
\end{equation}
This transformation law is determined in \cite{He:2014cra,Campiglia:2015qka} from a spacetime analysis. It can also be determined from \eqref{eq:CurrentDiv}. This suggests the following explicit form of $U_i$
\begin{equation}
\begin{split}
U_i[\t] \equiv \exp \left[ i Q_i \int d^d x \t(x) \CK_d ( m_i / \o_i , x_i ; x ) \right] \; .
\end{split}
\end{equation}
Note that this operator is not single-valued under the identification in \eqref{C-flat-theta}. Instead, we identify
\begin{equation}
\begin{split}
U_i[\t] \sim U_i[\t] e^{ i Q_i \t_0}  
\end{split}
\end{equation}
which follows from the property
\begin{equation}
\begin{split}
\int d^d x \CK_d ( m_i / \o_i , x_i ; x ) = 1 \; . 
\end{split}
\end{equation}
Including one such soft operator insertion for each of the operators $\CO_i$ appearing in the scattering amplitude, we find
\begin{equation}
\begin{split}\label{Sintabelian}
S_{\text{int}} [ \t , j ]  &= - i \int d^d x \t(x) \sum_i Q_i \CK_d ( m_i / \o_i , x_i ; x )  \; .
\end{split}
\end{equation}
Note that due to conservation of charge, the interaction action is single-valued under \eqref{C-flat-theta}, even though the operators $U_i$ are not. Using the definition of $j^a(x)$ in \eqref{Jains}, the action can be written 
\begin{equation}
\begin{split}
S_{\text{int}} [ \t , j ]  &=  i \int d^d x C_a (x)   j^a(x) \; , \qquad C = d \t  \; . \\
\end{split}
\end{equation}

Having fixed the form of $S_{\text{soft}}$ and $S_{\text{int}}$ we can evaluate the path integral in \eqref{IR-res-soft}. The propagator $P_{ab}$ will then be fixed by matching onto the RHS of \eqref{IR-res-soft}. For $U(1)$ gauge theories, the condition is
\begin{equation}
\begin{split}\label{tocalc}
 \int  [d \t ] e^{ - S_{\text{soft}} [\t] - S_{\text{int}} [\t , j ]  }   = e^{ - \G_\text{ph}}  = e^{ - \a  ( A_1 + 2\pi i A_2 )  } \; , 
\end{split}
\end{equation}
where
\begin{equation}
\begin{split}
A_1 =  \int  \frac{d^d x}{(2\pi)^d} \CJ^a(x) \CJ_a(x)  \;  , \qquad \a = \frac{e^2}{8\pi} \int_{\mu}^\L d\o \o^{d-3} \; .
\end{split}
\end{equation}
At this stage, we are unable to reproduce the imaginary term $A_2$ from this construction and consequently we postpone a discussion of this to future work. The path integral on the LHS is Gaussian and can be evaluated explicitly
\begin{equation}
\begin{split}
e^{ - \G_\text{ph}}  &= \int [ d \t ] \exp \left[ - \int d^d x d^d y \, (P^{-1})^{ab}(x-y)  C_a (x)  C_b(y) - i \int d^d x   C_a (x)  j^a(x)  \right] \; ,  \\
&=  \exp \left[  - \frac{1}{4} \int d^d x d^d y P_{ab} ( x - y ) j^a (x) j^b ( y ) \right]   \; . 
\end{split}
\end{equation}
Comparing this to \eqref{tocalc} and using \eqref{Jains}, we find that we can reproduce $A_1$ correctly if
\begin{equation}\label{eq:CurrentProp}
\begin{split}
P_{ab} ( x - y )  = \frac{16\a }{(2\pi)^d}  \int d^d w   \frac{ \CI_{ac} ( w - x ) }{ ( w - x )^2 } \frac{ \CI^c{}_b ( w - y ) }{ ( w - y )^2 } \; .
\end{split}
\end{equation}
This follows from the formal properties of the inverse shadow transform combined with \eqref{Jains}. 
To construct the action, we need to invert the propagator which satisfies
\begin{equation}
\begin{split}
\int d^d w ( P^{-1} )^{ac}  ( w - x ) P_{cb} ( w - y  ) = \d^a_b \d^{(d)}(x-y) \; .
\end{split}
\end{equation}
It follows that
\begin{equation}
\begin{split}
( P^{-1} )^{ab} ( x - y ) =  \frac{(2\pi)^d}{16 c_{1,1}^2 \a } \int d^d w \frac{ \CI^{ac} ( w - x ) }{ [ ( w - x )^2 ]^{d-1}  } \frac{ \CI_c{}^b ( w - y ) }{ [ ( w - y )^2]^{d-1}  } \; .
\end{split}
\end{equation}
The soft action is therefore
\begin{equation}
\begin{split}
S_{\text{soft}}[\t] &=   \frac{(2\pi)^d}{16 c_{1,1}^2 \a } \int   d^d w  \int  d^d x d^d y  \frac{ \CI^{ac} ( w - x )  \CI_c{}^b ( w - y )  }{ [ ( w - x )^2 ]^{d-1} [ ( w - y )^2]^{d-1}  } C_a(x) C_b(y)  \\
&= \frac{(2\pi)^d}{16 c_{1,1}^2 \a } \int d^d x {\wt C}^a(x) {\wt C}_a(x)   \; .
\end{split}
\end{equation}
Remarkably, although the pseudo-Goldstone action appears non-local when written in terms of the edge mode $C$, the shadow edge mode ${\wt C}$ behaves like a local degree of freedom. This may suggest that the ``appropriate'' set of local operators in celestial CFT$_d$ are the shadow transforms of bulk operators, and we hope to explore this possibility in future work. To conclude, we can also express the soft-interaction term using the shadow mode
\begin{equation}
\begin{split}\label{int-shadow-mode}
S_{\text{int}} [\t,j] =  i \int d^d x C_a (x)  j^a(x) =  \frac{i}{2c_{1,1}}  \int d^d x  {\wt C}_a (x) \CJ^a(x) \; .
\end{split}
\end{equation}
This local action for the shadow edge mode reproduces the infrared divergences in abelian gauge theory scattering amplitudes.

To end this discussion, we now consider scattering amplitudes with external soft photons. For this purpose, we are required to ``integrate in'' the soft photon operator $S = d \phi$. More precisely, we wish to determine an action $S_{\text{soft}}[\t,\phi]$ such that
\begin{equation}
\begin{split}\label{full-soft-act}
\int [ d \phi  ] e^{ - S_{\text{soft}}[\t,\phi] }  = e^{ - S_{\text{soft}}[\t]  } \; ,\qquad  \int [ d \phi  ] [d \t ] e^{ - S_{\text{soft}}[\t,\phi] - S_{\text{int}} [\t , j ]  }S_1 \cdots S_m = e^{ - \G}  \CJ_1 \cdots \CJ_m \; . 
\end{split}
\end{equation}
Note that the zero mode of $\phi$ does not appear in the physical operator $S$ and so is not part of the system described by $S_{\text{soft}}[\t,\phi]$. We therefore gauge the symmetry $\phi(x) \sim \phi(x) + c$.

The soft action can be determined as follows. The shadow transform of $S$ is an abelian current \eqref{eq:CurrentShadow}. This operator must generate the global $U(1)$ transformations $\theta \to \theta+ \varepsilon$. The usual Noether procedure then implies an off-diagonal coupling of the form
\begin{equation}
\begin{split}
S_{\text{soft}}[\t,\phi] &= S'_{\text{soft}}[\phi] - i \int d^d x C_a (x) J^a (x) = S'_{\text{soft}}[\phi] - \frac{i}{2c_{1,1}} \int d^d x {\wt C}_a (x) S^a (x) \; . 
\end{split}
\end{equation}
$S_{\text{soft}}'$ is fixed by imposing \eqref{full-soft-act}. The required term is Gaussian:
\begin{equation}
\begin{split}
S'_{\text{soft}}[\phi] = \frac{\a}{(2\pi)^d} \int d^d x S^a(x) S_a(x)  \; , \qquad S = d \phi \; . 
\end{split}
\end{equation}
The full soft action is therefore
\begin{equation}
\begin{split}\label{soft-action-abelian}
S_{\text{soft}}[\t,\phi] + S_{\text{int}} [\t,j]  = \frac{\a}{(2\pi)^d} \int d^d x S^a(x) S_a(x)  - \frac{i}{2c_{1,1}} \int d^d x  {\wt C}_a (x) [ S^a (x) - \CJ^a(x) ] \; ,
\end{split}
\end{equation}
with
\begin{equation}
\begin{split}
C = d \t \; , \qquad S = d \phi \; , \qquad  \phi(x) \sim \phi(x) + c \; , \qquad \t(x) \sim \t(x) + c \; . 
\end{split}
\end{equation}
This path integral correctly reproduces insertions of the soft photon operator
\begin{equation}
\begin{split}
\avg{ S_{a_1}(y_1) \cdots S_{a_m}(y_m) } = \int [ d \t ] [ d \phi ] e^{ - S_{\text{soft}}[\t,\phi] - S_{\text{int}} [\t,j]  } \p_{a_1} \phi(y_1) \cdots \p_{a_m} \phi (y_m) \; . 
\end{split}
\end{equation}
The integral over $\t(x)$ gives a Dirac delta function $\d ( S - \CJ )$ which then localizes the integral over $\phi$. We find
\begin{equation}
\begin{split}
\avg{ S_{a_1}(y_1) \cdots S_{a_m}(y_m) } =   \exp \left[ - \frac{\a}{(2\pi)^d} \int d^d x \CJ^a(x) \CJ_a(x)   \right] \CJ_{a_1}(y_1) \cdots \CJ_{a_m}(y_m)
\end{split}
\end{equation}
which is (almost\footnote{As mentioned previously, we do not yet understand how the imaginary part of $\G_{\text{ph}}$ is reproduced by the soft action and we leave this to future work.}) the expected result \eqref{IR-res-soft-1}!

\subsubsection{Magnetic Charges and Winding Modes in $d=2$}
\label{sec:magcharge2d}

We have so far demonstrated that the action \eqref{soft-action-abelian} reproduces all infrared divergences \emph{and} soft theorems under the assumption that all the particles in the scattering process are purely electrically charged. In this section, we generalize the result to include dyonic scattering states in four dimensions. In order to simplify the discussion, we will assume that all hard particles are massless.

In $d=2$  the shadow transform of a flat connection localizes, and the Goldstone mode is proportional to its shadow:
\begin{equation}
\begin{split}
\wt{\p_a \t}(x) = \int d^2 x \frac{ \CI_{ab} ( x - y ) }{ ( x - y )^2 } \p^b \t ( y ) =  2\pi \p_a \t(x) \; . 
\end{split}
\end{equation}
The action \eqref{soft-action-abelian} then reduces to
\begin{equation}
\begin{split}
S_{\text{soft}}[\t,\phi] &=  \frac{1}{4\pi^2} \int   [ \a d \phi  \w \star d \phi  - \pi  i \, d \t  \w \star d \phi  ] \; .
\end{split}
\end{equation}
The soft correlators for dyonic charges are given by \eqref{Sains-dyon} and \eqref{eq:MagDiv} : 
\begin{equation}
\begin{split}
\avg{ S_{a_1}(y_1) \cdots S_{a_m}(y_m) }  &=  [ \CJ_{a_1} (y_1) + {\wt \CJ}_{a_1} (y_1) ]  \cdots [ \CJ_{a_m} (y_m) + {\wt \CJ}_{a_m} (y_m) ] \\
&\qquad \qquad \qquad \qquad \times \exp \left[ - \frac{\a}{(2\pi)^d} \int d^d x  \left[ \CJ_a(x) + {\wt \CJ}_a(x) \right]^2   \right] 
\end{split}
\end{equation}
where
\begin{equation}
  \CJ_a (x) = \p_a \sum_i Q_i \log [ - {\hat p}_i \cdot {\hat q}(x) ] \; , \qquad 
{\wt \CJ}_a (x) \equiv \frac{2\pi}{e^2} \e_{ab} \p^b \sum P_i  \log[ -  {\hat p}_i \cdot {\hat q}(x) ] \; . 
\end{equation}
It is easy to see that the soft action \eqref{soft-action-abelian} with an additional source
\begin{equation}
\begin{split}
S_{\text{soft}}[\t,\phi] + S_{\text{int}} [\t,j]  = \frac{\a}{(2\pi)^2} \int d^2 x  S^a(x) S_a(x) - \frac{i}{4\pi} \int d^2 x C_a (x) [ S^a  (x) - \CJ^a(x)-{\wt \CJ}_a(x) ] \; 
\end{split}
\end{equation}
correctly reproduces these correlation functions. The new interaction term $\wt{\CJ}^a C_a(x) $ can be interpreted as a background source for the topological winding current $\epsilon_{ab}\p^b \theta$, just as the original term $\p_a \theta\CJ^a$ may be interpreted as a background source for the momentum current $\p_a \theta$. In this way, the electric and magnetic charges of the abelian gauge theory are mapped to the momentum and winding charges of the vertex operators in the compact boson theory. The irrational dyonic spectrum in the presence of the $\vartheta$ term is simply reproduced by a background Kalb-Ramond field. The coupling $iB_{\theta \phi}d\theta\wedge d\phi$ simply changes the spectrum of allowed charges in the soft sector, but does not alter the dynamics. The final soft action takes the form
\begin{equation}
\begin{split}
S_{\text{soft}}[\t,\phi] &=  - \frac{1}{4\pi^2} \int \left[ \a d \phi  \w \star d \phi  - \pi  i \, d \t   \w  \star \left(  d \phi + \frac{\tau_1}{\tau_2} \star d \phi  \right)  \right] \; .
\end{split}
\end{equation}

 The structure of this term can actually be determined using the bulk $\vt$-term
\begin{equation}
\begin{split}
 i S_{\text{bulk}} \ni   \frac{i \vt}{8\pi^2 m_0 } \int d(A\wedge F) \; .
\end{split}
\end{equation}
This is a boundary term and its contribution to $S_{\text{soft}}$ can be determined by integrating over $\ci$
\begin{equation}
\begin{split}
S_{\text{soft}} \ni  \int_{\ci}  A  \w F  \sim  \int d \t \w d \phi \; .
\end{split}
\end{equation}

\subsection{Non-abelian Gauge Theory}

The symmetry-based part of the previous discussion generalizes straightforwardly to non-abelian gauge theories. The group $\mathcal{G}$ of maps $S^d\to G$ is spontaneously broken on the perturbative vacuum, and the corresponding order parameter is a non-abelian boundary Wilson line
\begin{equation}
\begin{split}
W(x) = \CP \left( \exp \int_{x_0}^{x} C \right) \; , \qquad C = A |_{\p \ci} \;  . 
\end{split}
\end{equation}
Under large gauge transformations
\begin{equation}
\begin{split}
W(x) \to g(x) W(x) g(x_0)^{-1} \; , \qquad g(x) \in G \; . 
\end{split}
\end{equation}
In the quantum theory, the Wilson line has a non-vanishing vacuum expectation value $\avg{ W(x) } \neq 0$ when the gauge theory is in the Coulomb phase. Global gauge transformations leave the trace of the order parameter invariant, so the large gauge symmetry is broken down to global $G$ symmetry and the corresponding Goldstone mode is the flat connection $C=U d U^{-1}$. In the absence of explicit breaking by the infrared regulator, the Goldstone action must be invariant under the $d$-dimensional local invariance group $\mathcal{G}$ and is therefore a gauge theory without a quotient by the orbit of the gauge group. The flatness condition is imposed by a non-abelian $BF$ term of the form $i \int \tr{ B \w F }$. The symmetry breaking term is the non-abelian generalization of the non-local vector mass term \eqref{eq:AbelianBreaking}
\begin{equation}
\begin{split}\label{ymaction}
S_{\text{soft}}[U] = \int d^d x d^d y (P^{-1})^{ab}(x-y) \tr{ C_a(x) C_b(y) } \; , \qquad C = U d U^{-1} \; .
\end{split}
\end{equation}
In $d=2$, there is another possible symmetry breaking term that would give an action to the pseudo-Goldstones, namely the WZW term, and we expect such a term to follow from the addition of a non-abelian $\vartheta$-term in the bulk Lagrangian. Higher-dimensional $\vartheta$-terms similarly correspond to higher-dimensional WZW terms.

The interaction term $S_{\text{int}}$ is also determined by symmetries. To describe this, we define the group element $U(x) = e^{ - \t(x) } $ where $\t(x)\equiv \t^I(x) T^I  \in \mfg$. We define the operator $U_i[\t]$ as
\begin{equation}\label{eq:nonAbelianOp}
\begin{split}
U_i[\t] = R_i \left( \exp \left[  \int d^d x \t(x) \CK_d ( m_i / \o_i , x_i ; x )\right] \right) \; .
\end{split}
\end{equation}
The interaction term $e^{-S_{\text{int}}}$ is obtained as a product of these operators, one for each of the asymptotic states in the amplitude. 

The discussion so far is completely general and used only the symmetries of the problem. The main complication arises in the ``interdimensional effective field theory'' matching condition. In order to determine the propagator entering in \eqref{ymaction}, we need to match on to the non-abelian infrared divergence, which is not known exactly and is severely more complicated than the abelian case. Similarly, the soft gluon theorem receives perturbative corrections so it is not as simple to ``integrate in'' the external soft operator. To make progress we have to revert to perturbation theory, keeping in mind that the soft dynamics is ultimately non-perturbative. It is well known that the model defined by \eqref{ymaction} with a local propagator (the principal chiral model in two dimensions) is also strongly coupled, and that perturbation theory about the free point does not accurately capture the dynamics and symmetry restoration (Coleman-Wagner theorem). Ultimately, one hopes to apply non-perturbative knowledge of the strong coupling behavior in the principal chiral model to the much less understood strong coupling problem in Yang Mills. This is left to future work. Upon linearizing the problem
\begin{equation}
U(x)=1-\t(x)+ \dots
\end{equation}
the action \eqref{ymaction} describes a theory of (free) scalars
\begin{equation}
\begin{split}
S_{\text{soft}}[\t] = \int d^d x d^d y (P^{-1})^{ab}(x-y) \tr{ \p_a \t (x) \p_b \t(y) } \; .
\end{split}
\end{equation}
This linearized action (with a local propagator) was recently used in \cite{Magnea:2021fvy} to match onto and reproduce low orders in four-dimensional Yang-Mills perturbation theory. It seems natural to conjecture that the non-linear completion of that model
\begin{equation}
S_{\text{soft}} [\t,\phi] =  \int d^dx \, \tr{ \frac{\a}{(2\pi)^d} S^a(x) S_a(x) - \frac{i}{2c_{1,1}}  S_a(x) {\wt C}^a(x) } 
\end{equation}
describes higher orders in perturbation theory. The interaction term would then be obtained through insertions of the operators \eqref{eq:nonAbelianOp}. This path integral cannot be performed exactly, but could be compared at any order in perturbation theory. We hope to investigate this possibility in future work.

\subsection{Gravity: Supertranslation Mode}

The asymptotic symmetry group for gravitational theories is the BMS group \cite{Bondi:1962px,Sachs:1962wk,He:2014laa,Kapec:2015vwa}. The implementation of this symmetry in the celestial CFT is slightly complicated given that one is trying to realize higher-dimensional spacetime symmetries as internal symmetries in a lower-dimensional model. The BMS group consists of $SO(1,d+1)$ Lorentz transformations as well as supertranslations, which are an infinite-dimensional extension of the $(d+2)$ translation group.\footnote{There is evidence that the Lorentz group should be viewed as the unbroken subgroup of diffeomorphisms of the celestial sphere, although a proper asymptotic analysis has not yet been performed and the question remains open.} The BMS symmetries are spontaneously broken down to the Poincar\'e group by the perturbative vacuum. The action for the corresponding Goldstone modes is expected to be BMS invariant in the absence of explicit breaking.

The standard asymptotic analysis identifies the supertranslation Goldstone mode as the leading traceless boundary metric fluctuation $C_{ab} = ( h_{ab} - \frac{1}{d} \d_{ab} h) |_{\p \ci}$ which satisfies the flatness condition\footnote{As in gauge theory, \eqref{Cabflatness} is derived from the vanishing of the asymptotic field strength (in this case, the Weyl tensor component $C_{arbc} |_{\p\ci}$).}
\begin{equation}
\begin{split}\label{Cabflatness}
\o_{abc} \equiv \p_{[b} C_{c]a}  - \frac{1}{d-1} \d_{a[b} \p^d C_{c]d} = 0 \; .
\end{split}
\end{equation}
Supertranslations act non-linearly on the Goldstone mode
\begin{equation}
\begin{split}\label{sttransform}
C_{ab} (x) \to C_{ab}(x) + 2  \left( \p_a \p_b - \frac{1}{d} \d_{ab} \p^2 \right) f(x) \; 
\end{split}
\end{equation}
but leave $\omega_{abc}$ invariant. Here $f(x)$ is any function on $S^d$ and satisfies no periodicity condition. The infinite-dimensional supertranslation symmetry is spotaneously broken to a finite-dimensional translation symmetry which is generated by functions of the form
\begin{equation}
\begin{split}\label{translations}
f_{tr}(x) = \chi^0  ( 1+x^2 ) - 2 \chi^a x_a - \chi^{d+1}  ( 1- x^2 ) 
\end{split}
\end{equation}
which leave $C_{ab}(x)$ invariant. 

Following the gauge theory construction, our first step is to find an action which is invariant under the full supertranslation symmetry. The field strength $\o_{abc}$ is invariant under supertranslations \eqref{sttransform} so the simplest model is a ``supertranslation BF theory''
\begin{equation}
\begin{split}
S = i \int d^d x B^{abc} \o_{abc} \; . 
\end{split}
\end{equation}
Integrating out the $B$ field imposes the flatness condition $\o_{abc} = 0 $ so we can write
\begin{equation}
\begin{split}\label{Cidentification-gr}
C_{ab}(x) = 2 \left( \p_a \p_b - \frac{1}{d} \d_{ab} \p^2 \right) C(x) \; , \qquad C(x) \sim C(x) + f_{tr}(x) \; .
\end{split}
\end{equation}
The identification is required because the global translation symmetry is unbroken on the vacuum. Under supertranslations
\begin{equation}
\begin{split}\label{Csupertranslation}
C(x) ~~ \stackrel{\CG}{\longrightarrow} ~~  C(x) + f(x) \;  . 
\end{split}
\end{equation}
The leading  symmetry breaking term gives an action to pure supertranslations and takes the form
\begin{equation}
S_{\text{soft}}[C] =  \int d^d x d^d y ( P^{-1} )^{ab,cd} ( x - y ) C_{ab}(x) C_{cd} ( y ) \; .
\end{equation}
The interaction term follows from the symmetry transformation properties of the plane wave\footnote{For the considerations of this section, we do not work in the Mellin-transformed basis. } creation and annihilation operators \cite{Campiglia:2015kxa}
\begin{equation}
\begin{split}
\CO_i ( \o_i , x_i ) ~~ \stackrel{\CG}{\longrightarrow}~~ \exp \left[ \frac{i}{2} m_i \int d^d x f(x) \CK_{d+1} ( m_i / \o_i , x_i ; x ) \right] \CO_i ( \o_i , x_i ) \; .
\end{split}
\end{equation}
In particular, under the unbroken symmetry transformations (translations) generated by \eqref{translations}, these operators transform as  plane waves
\begin{equation}
\begin{split}
\CO_i ( \o_i , x_i )  ~~ \stackrel{G}{\longrightarrow} ~~  e^{ - i p_i \cdot \chi } \CO_i ( \o_i , x_i ) \; .
\end{split}
\end{equation}
It then follows from \eqref{Csupertranslation} that the operator serving as a classical source for $C(x)$ is 
\begin{equation}
\begin{split}
U_i[C] \equiv \exp \left[ \frac{i}{2} m_i \int d^d x C(x) \CK_{d+1} ( m_i / \o_i , x_i ; x ) \right] \; , \qquad U_i[C]  \sim U_i[C] e^{ - i p_i \cdot \chi } \; . 
\end{split}
\end{equation}
The soft-interaction term therefore takes the form
\begin{equation}
\begin{split}
S_{\text{int}} [ C , j ]  &= - \frac{i}{2} \int d^d x C(x) \sum_i m_i \CK_{d+1}  ( m_i / \o_i , x_i ; x ) \; .
\end{split}
\end{equation}
Using the definition of $j_{ab}(x)$ in \eqref{jabdef-gr}, we can rewrite this as
\begin{equation}
\begin{split}
S_{\text{int}} [ C , j ]  &= - \frac{i}{4} \int d^d x C^{ab} (x) j_{ab}(x)  \; .
\end{split}
\end{equation}
Note that due to momentum conservation, the interaction action is invariant under \eqref{Cidentification-gr} even though the operators $U_i$ are not.

Having fixed both the soft action and the interaction term, we can now determine the propagator $P_{ab,cd}$ by imposing the constraint \eqref{IR-res-soft}. For gravitational theories this reads
\begin{equation}
\begin{split}\label{gr-constraint-soft}
\int[ dC] e^{ - S_{\text{soft}}[C] - S_{\text{int}} [ C , j ] } = e^{ - \G_{\text{gr}} } = e^{ - \a_{\text{gr}} ( A_1^{\text{gr}} + 2\pi i A_2^{\text{gr}} )  }  \; ,
\end{split}
\end{equation}
where
\begin{equation}
\begin{split}
A_1^{\text{gr}} =  \int  \frac{d^d x}{(2\pi)^d}  \CJ^{ab}(x) \CJ_{ab}(x)   \;  , \qquad \a_{\text{gr}} = \frac{\kappa^2}{32\pi} \int_{\mu}^\L d\o \o^{d-3}\; .
\end{split}
\end{equation}
As in the abelian case, we leave a discussion of the imaginary term $A_2^{\text{gr}}$ to future work. The soft path integral is Gaussian and we can evaluate it explicitly. The condition \eqref{gr-constraint-soft} then implies
\begin{equation}
\begin{split}\label{required-property-gr}
\frac{1}{64} \int d^d x d^d y P_{ab,cd}(x-y) j^{ab}(x) j^{cd}(y) =  \a_{\text{gr}}  \int \frac{d^dx}{(2\pi)^d} \CJ^{ab}(x) \CJ_{ab}(x)  \; . 
\end{split}
\end{equation}
The propagator $P$ satisfies 
\begin{equation}
\begin{split}
\int d^d w P_{ab,ef} ( w - y  )( P^{-1} )^{ab,cd}  ( w - x )  = \d^{\{c}_{\{e}\d^{d\}}_{f\}} \d^{(d)}(x-y) \; ,
\end{split}
\end{equation}
where the curly braces denote traceless symmetrization. Using the relationship between $j_{ab}$ and $\CJ_{ab}$ in \eqref{jabdef-gr}, we determine the propagator to be
\begin{equation}
\begin{split}
P_{ab,cd}(x-y) = \frac{ 1024 \a_{\text{gr}} }{ (2\pi)^d }   \int d^dw  \frac{ \CI_{e\{a}(x-w) \CI_{b\}f} ( x - w ) }{  ( x - w )^2 }  \frac{ \CI^e{}_{\{c} (y-w) \CI^f{}_{d\}} ( y - w ) }{  ( y - w )^2 } \; . 
\end{split}
\end{equation}
Consequently,
\begin{equation}
\begin{split}
(P^{-1})^{ab,cd}(x-y) =  \frac{ (2\pi)^d }{ 1024c_{1,2}^2 \a_{\text{gr}}  }   \int d^dw  \frac{ \CI^{e\{a}(x-w) \CI^{b\}f} ( x - w ) }{ [ ( x - w )^2 ]^{d-1} }  \frac{ \CI_e{}^{\{c} (y-w) \CI_f{}^{d\}} ( y - w ) }{ [ ( y - w )^2 ]^{d-1}  } \; . 
\end{split}
\end{equation}
The gravitational soft action is therefore
\begin{equation}
\begin{split}\label{softaction-gr}
S_{\text{soft}}[C] &= \frac{ (2\pi)^d }{ 1024 c_{1,2}^2  \a_{\text{gr}}}   \int d^d x {\wt C}_{ab}(x)  {\wt C}^{ab}(x)  \; .
\end{split}
\end{equation}
As in gauge theory, the shadow gravity mode behaves like a local degree of freedom. This provides further evidence that the set of local operators in the celestial CFT are shadow transforms of bulk operators. In terms of the shadow mode, the soft interaction term takes the form
\begin{equation}
\begin{split}
S_{\text{int}} [ C , j ]  &= - \frac{i}{16c_{1,2}} \int d^d x {\wt C}^{ab} (x)   \CJ_{ab} (x)   \; .
\end{split}
\end{equation}

To finish this discussion, we `integrate in' the soft graviton mode $N_{ab}$ following the same procedure as in the abelian case. The resultant soft action is
\begin{equation}
\begin{split}
S_{\text{soft}}[C,N] = \frac{\a_{\text{gr}}}{(2\pi)^d} \int d^d x N_{ab}(x) N^{ab}(x) + \frac{i}{16 c_{1,2}} \int d^d x N_{ab}(x) {\wt C}^{ab}(x)  \; , 
\end{split}
\end{equation}
where $N_{ab}(x) = 2(\p_a \p_b - \frac{1}{d} \d_{ab} \p^2 ) N(x)$. Insertions of the soft graviton operator are then given by
\begin{equation}
\begin{split}
\avg{ N_{a_1b_1}(y_1) \cdots N_{a_m b_m}(y_m) } = \int [ d N ] [ d C ] e^{ - S_{\text{soft}}[C,N] - S_{\text{int}}[C,j] } N_{a_1b_1}(y_1) \cdots N_{a_m b_m}(y_m)  \; .
\end{split}
\end{equation}
The integral over $C(x)$ gives a Dirac delta functional $\d(N-\CJ)$ which then localizes the integral over $N$. The result is
\begin{equation}
\begin{split}
\avg{ N_{a_1b_1}(y_1) \cdots N_{a_m b_m}(y_m) } =  \CJ_{a_1b_1}(y_1) \cdots \CJ_{a_m b_m}(y_m) \exp \left[ - \frac{\a_{\text{gr}}}{(2\pi)^d} \int d^d x \CJ_{ab}(x) \CJ^{ab}(x)  \right] \; 
\end{split}
\end{equation}
which is expected from \eqref{IR-res-soft-1}.

\section{Conclusions and Future Work}\label{sec:Conclusion}
We have demonstrated that symmetry principles are sufficient in order to describe soft dynamics in gauge theory and gravity. In any dimension, the exponentiated abelian soft exchange arises straightforwardly from a pattern of spontaneous and explicit symmetry breaking. The effective dynamics of the long-wavelength Goldstone bosons associated to gauge transformations with non-compact support are effectively lower-dimensional, and their interactions are captured by lower-dimensional gauge theories (without a gauge quotient) deformed by symmetry breaking mass terms.  Although the effective action expressed in terms of the Goldstone edge mode appears non-local, the shadow Goldstone has local dynamics and reproduces the effects of soft exchange in the bulk model.

In addition to providing a description of abelian soft exchange in the celestial CFT formalism, this paper raises two unanswered questions which we hope to explore in future work. The first question regards the celestial CFT interpretation of the phase of the exponentiated soft exchange. The imaginary part of $\G$ differs significantly from its real part since it only receives contributions from pairs of incoming particles and pairs of outgoing particles. It is not clear to us how to reproduce this qualitative structure from a soft action.

The second problem that requires further investigation is the non-abelian soft exchange amplitude. The long-wavelength gluon and the associated pseudo-Goldstone boson are strongly coupled in $d=2$, and both path integrals can only be approximated perturbatively. Much more is known and proven in the 2D non-linear sigma model than in 4D Yang-Mills, and ultimately one would like to connect the well understood strong coupling behavior of the 2D model to confinement in the 4D gauge theory. Since it is precisely the long-wavelength fluctuations of the chromo-electric field that are strongly coupled, this connection with the lower-dimensional pseudo-Goldstone seems like a promising way to proceed beyond perturbation theory. We hope to pursue this direction in future work. We also feel that the soft theorems for non-abelian chromodyons and magnetic branes conjectured in Section \ref{sec:MagSoft} deserve further exploration.

\subsection*{Acknowledgments}
We would like to thank Amr Ahmadain, Elizabeth Himwich, Sruthi Narayan, Monica Pate, Ana-Maria Raclariu and Andrew Strominger for many useful conversations. DK gratefully acknowledges support from U.S. Department of Energy grant DE-SC0009988 and the Adler Family Fund. PM gratefully acknowledges support from U.S. Department of Energy grant DE-SC0009988, the Infosys Fellowship and STFC consolidated grants ST/P000681/1, ST/T000694/1.

\appendix

\section{Derivation of Current Ward Identities}
\label{app:wardidderivation}

In this section, we derive the current Ward identities \eqref{eq:CurrentDiv}, \eqref{eq:CurrentDiv-ym} and \eqref{eq:CurrentDivGR}. We start with the abelian Ward identity. To prove the result, we need to show that
\begin{equation}
\begin{split}\label{abelian-toprove}
\p^a j_a(x) =  \sum_i Q_i \CK_d ( m_i / \o_i , x_i ; x )
\end{split}
\end{equation}
where
\begin{equation}
\begin{split}\label{Jadef-app-1}
j_a(x) = \lim_{\D \to 1 }  \frac{1}{2c_{\D,1}} \int d^d y \frac{\CI_{ab}(x-y)}{ [ ( x - y )^2 ]^{d-\D} } \CJ^b(y) \; ,  \qquad \CJ_a (x) = \p_a \sum_i Q_i \log [ - {\hat p}_i \cdot {\hat q}(x) ] \; . 
\end{split}
\end{equation}
Note that we have regulated the shadow integral so that it is well-defined in any dimension. Using this definition, the LHS of \eqref{abelian-toprove} is 
\begin{equation}
\begin{split}
\p^a j_a(x) =   \lim_{\D \to 1 } \frac{1}{2c_{\D,1}} \int d^d y \p^a \frac{\CI_{ab}(x-y)}{ [ ( x - y )^2 ]^{d-\D} } \CJ^b(y)  \; .
\end{split}
\end{equation}
Using the identity
\begin{equation}
\begin{split}
\p^a \frac{\CI_{ab}(x)}{ (x^2)^\l }  = \frac{1}{\l} ( d - 1 - \l ) \p_b \frac{1}{(x^2)^\l} 
\end{split}
\end{equation}
we find
\begin{equation}
\begin{split}
\p^a j_a(x) =  \lim_{\D \to 1 } \frac{\D-1}{2c_{\D,1}(d-\D)} \int d^d y  \frac{1}{ [ ( x - y )^2 ]^{d-\D} }  \p^a \CJ_a (y) \; .
\end{split}
\end{equation}
Using the definition \eqref{Jadef-app-1}, we find
\begin{equation}
\begin{split}
\p^a \CJ_a (x) =  \sum_i Q_i \left( \frac{d-2}{ [ - {\hat p}_i \cdot {\hat q}(x) ] }  +  \frac{ z_i^2 }{ [ - {\hat p}_i \cdot {\hat q}(x) ]^2 }  \right) , \qquad z_i = m_i / \o_i 
\end{split}
\end{equation}
which implies 
\begin{equation}
\begin{split}\label{currentdiv-eval-app}
\p^a j_a(x) =  \lim_{\D \to 1 } \frac{\D-1}{2c_{\D,1}(d-\D)}  \sum_i Q_i [ 2 ( d - 2 ) I_{\D,1} ( z_i , x - x_i )  +  4 z_i^2 I_{\D,2} ( z_i , x - x_i ) ]
\end{split}
\end{equation}
where
\begin{equation}
\begin{split}\label{intab-def}
I_{a,b}(z,x) = \int d^d y \frac{1}{ [ y^2 ]^{d-a} [ ( y + x )^2 + z^2 ]^b } \; . 
\end{split}
\end{equation}
To evaluate this integral we use the Feynman-Schwinger parameterization
\begin{equation}
\begin{split}
\label{FSeq}
\frac{1}{X_1^{a_1} \cdots X_n^{a_n} } &= \frac{\G ( \sum a_i ) }{ \G ( a_1 ) \cdots \G ( a_n ) }\int_0^1 \frac{dt_1}{t_1} \cdots \frac{dt_n}{t_n}  \d \big( 1 - \sum_i t_i \big)  \frac{t_1^{a_1} \cdots t_n^{a_n} }{ ( \sum t_i X_i  )^{ \sum a_i } } 
\end{split}
\end{equation}
which implies
\begin{equation}
\begin{split}
I_{a,b}(z,x) &= \frac{\G(d-a+b)}{\G(d-a)\G(b)} \int_0^1 dt (1-t)^{d-a-1} t^{b-1} \int d^d y \frac{1}{ [ (1 - t ) y^2 + t ( y + x )^2 + t z^2 ]^{d-a+b} }  \\
&= \frac{\G(d-a+b)}{\G(d-a)\G(b)} \int_0^1 dt (1-t)^{d-a-1} t^{b-1} \int d^d y \frac{1}{ [ ( y + t x )^2 + t ( 1 - t ) x^2  + t z^2 ]^{d-a+b} } \; . 
\end{split}
\end{equation}
We can now shift $y \to y - t x$ and then move to spherical coordinates to evaluate the $y$ integral,
\begin{equation}
\begin{split}
\int d^d y \frac{1}{ [ ( y + t x )^2 + t ( 1 - t ) x^2  + t z^2 ]^{d-a+b} } &= \frac{2\pi^{d/2}}{\G(d/2)} \int_0^\infty dr \frac{r^{d-1} }{ [ r^2 + t ( 1 - t ) x^2  + t z^2 ]^{d-a+b} } \\
&=  \frac{ \pi^{d/2}\G ( \frac{d}{2} - a + b )   }{   \G (  d - a + b ) } [ t ( 1 - t ) x^2  + t z^2 ]^{a - b - \frac{d}{2} } \; . 
\end{split}
\end{equation}
The integral takes the form
\begin{equation}
\begin{split}
I_{a,b}(z,x) &= \frac{ \pi^{d/2}\G ( \frac{d}{2} - a + b )   }{ \G(d-a)\G(b) }  \int_0^1 dt t^{ a - \frac{d}{2} - 1 }   (1-t)^{d-a-1} [ ( 1 - t ) x^2  +  z^2 ]^{a - b - \frac{d}{2} } \; . 
\end{split}
\end{equation}
The integral over $t$ is related to the integral representation of the Gauss hypergeometric function
\begin{equation}
\begin{split}
\label{hgf}
\,_2 F_1(a,b,c;z)=\frac{\G(c)}{\G(b)\G(c-b)} \int_0^1 dt t^{b-1}(1-t)^{c-b-1}(1-tz)^{-a} \; . 
\end{split}
\end{equation}
Using this, we  have
\begin{equation}
\begin{split}\label{intab-eval}
I_{a,b}(z,x) &= \frac{ \pi^{d/2}\G ( \frac{d}{2} - a + b )  \G ( a-\frac{d}{2} )  }{ \G(b) \G ( d / 2 ) }  [ x^2  +  z^2    ]^{a - b - \frac{d}{2} } \,_2 F_1 \left( \frac{d}{2}-a+b , a - \frac{d}{2} ; \frac{d}{2} ; \frac{ x^2 }{ x^2 + z^2 } \right) \; . 
\end{split}
\end{equation}
Plugging this into \eqref{currentdiv-eval-app} and taking $\D \to 1$, we find
\begin{equation}
\begin{split}
\p^a j_a(x) =   \sum_i Q_i \CK_d ( m_i / \o_i , x_i ; x ) \; . 
\end{split}
\end{equation}
The derivation of the non-abelian current Ward identity \eqref{eq:CurrentDiv-ym} immediately follows from the above calculation as well. We now turn to the gravitational Ward identity \eqref{eq:CurrentDivGR}. To prove this, we need to show that
\begin{equation}
\begin{split}
\p^a \p^b j_{ab} (x) = \sum_i m_i \CK_{d+1} ( m_i / \o_i , x_i ; x ) \; , 
\end{split}
\end{equation}
where
\begin{equation}
\begin{split}\label{pab-def-reg}
j_{ab}(x) = - \lim_{\D \to 1} \frac{1}{4c_{\D,2}}  \int d^d y \frac{ \CI_{ac}(x-y) \CI_{bd} ( x - y ) }{[( x - y )^2]^{d-\D} } \CJ^{cd} (y) 
\end{split}
\end{equation}
and
\begin{equation}
\begin{split}\label{app-Jabdef}
\CJ_{ab}(x) &\equiv - \left( \p_a \p_b - \frac{1}{d} \d_{ab} \p^2 \right) \sum_i \o_i [ - {\hat p}_i \cdot {\hat q}(x) ]  \log [ - {\hat p}_i \cdot {\hat q}(x) ] \;  .
\end{split}
\end{equation}
As in the abelian case, we have regulated the shadow integral to make it well defined.

Using the identity
\begin{equation}
\begin{split}
\p^a \p^b \frac{ \CI_{a\{c}(x) \CI_{d\}b} (x) }{ ( x^2 )^\l } = \frac{(\l-d )( \l - d + 1 )}{ \l ( \l + 1 ) }   \p_{\{c}  \p_{d\}} \frac{1}{ ( x^2 )^\l } 
\end{split}
\end{equation}
we find
\begin{equation}
\begin{split}\label{ddjgr}
\p^a \p^b j_{ab} (x) &=-  \lim_{\D \to 1} \frac{1}{4 c_{\D,2} } \frac{\D(\D-1)}{ (d-\D) ( d-\D + 1 ) }   \int d^d y   \frac{1}{[( x - y )^2]^{d-\D} } \p_c \p_d  \CJ^{cd} (y) \; . 
\end{split}
\end{equation}
To simplify this further, we use
\begin{equation}
\begin{split}
\p^a \p^b \CJ_{ab}(x) = - \frac{d-1}{d} \sum_{i=1}^n \o_i \left( \frac{d(d-2)}{[ - {\hat p}_i \cdot {\hat q}(x) ]} + \frac{2 ( d - 2 )z_i^2}{[ - {\hat p}_i \cdot {\hat q}(x) ]^2} + \frac{2z_i^4}{[ - {\hat p}_i \cdot {\hat q}(x) ] ^3} \right) \; . 
\end{split}
\end{equation}
It follows that
\begin{equation}
\begin{split}
\p^a \p^b j_{ab} (x) &=  \lim_{\D \to 1}  \frac{(d-1)\D(\D-1)}{ 4 c_{\D,2} d (d-\D) ( d-\D + 1 ) } \sum_i \o_i  [ 2 d ( d - 2 ) I_{\D,1} (z_i , x - x_i )  \\
& \qquad \qquad \qquad \qquad \qquad + 8 ( d - 2 ) z_i^2 I_{\D,2} (z_i , x - x_i ) + 16 z_i^4  I_{\D,3} (z_i , x - x_i )  ] \; .
\end{split}
\end{equation}
Finally, we use \eqref{intab-eval} and take the $\D \to 1$ limit to find
\begin{equation}
\begin{split}
\p^a \p^b j_{ab} (x) &=  \sum_{i=1}^n m_i \CK_{d+1} ( m_i / \o_i , x_i ; x ) \; . 
\end{split}
\end{equation}

\bibliographystyle{apsrev4-1long}
\bibliography{Bib.bib}

\begin{thebibliography}{10}%
\makeatletter
\providecommand \@ifxundefined [1]{%
 \ifx #1\undefined \expandafter \@firstoftwo
 \else \expandafter \@secondoftwo
\fi
}%
\providecommand \@ifnum [1]{%
 \ifnum #1\expandafter \@firstoftwo
 \else \expandafter \@secondoftwo
\fi
}%
\providecommand \enquote [1]{``#1''}%
\providecommand \bibnamefont  [1]{#1}%
\providecommand \bibfnamefont [1]{#1}%
\providecommand \citenamefont [1]{#1}%
\providecommand\href[0]{\@sanitize\@href}%
\providecommand\@href[1]{\endgroup\@@startlink{#1}\endgroup\@@href}%
\providecommand\@@href[1]{#1\@@endlink}%
\providecommand \@sanitize [0]{\begingroup\catcode`\&12\catcode`\#12\relax}%
\@ifxundefined \pdfoutput {\@firstoftwo}{%
 \@ifnum{\z@=\pdfoutput}{\@firstoftwo}{\@secondoftwo}%
}{%
 \providecommand\@@startlink[1]{\leavevmode\special{html:<a href="#1">}}%
 \providecommand\@@endlink[0]{\special{html:</a>}}%
}{%
 \providecommand\@@startlink[1]{%
  \leavevmode
  \pdfstartlink
   attr{/Border[0 0 1 ]/H/I/C[0 1 1]}%
   user{/Subtype/Link/A<</Type/Action/S/URI/URI(#1)>>}%
  \relax
 }%
 \providecommand\@@endlink[0]{\pdfendlink}%
}%
\providecommand \url  [0]{\begingroup\@sanitize \@url }%
\providecommand \@url [1]{\endgroup\@href {#1}{\urlprefix}}%
\providecommand \urlprefix [0]{URL }%
\providecommand \Eprint[0]{\href }%
\@ifxundefined \urlstyle {%
  \providecommand \doi [1]{doi:\discretionary{}{}{}#1}%
}{%
  \providecommand \doi [0]{doi:\discretionary{}{}{}\begingroup
  \urlstyle{rm}\Url }%
}%
\providecommand \doibase [0]{http://dx.doi.org/}%
\providecommand \Doi[1]{\href{\doibase#1}}%
\providecommand \bibAnnote [3]{%
  \BibitemShut{#1}%
  \begin{quotation}\noindent
    \textsc{Key:}\ #2\\\textsc{Annotation:}\ #3%
  \end{quotation}%
}%
\providecommand \bibAnnoteFile [2]{%
  \IfFileExists{#2}{\bibAnnote {#1} {#2} {\input{#2}}}{}%
}%
\providecommand \typeout [0]{\immediate \write \m@ne }%
\providecommand \selectlanguage [0]{\@gobble}%
\providecommand \bibinfo [0]{\@secondoftwo}%
\providecommand \bibfield [0]{\@secondoftwo}%
\providecommand \translation [1]{[#1]}%
\providecommand \BibitemOpen[0]{}%
\providecommand \bibitemStop [0]{}%
\providecommand \bibitemNoStop [0]{.\EOS\space}%
\providecommand \EOS [0]{\spacefactor3000\relax}%
\providecommand \BibitemShut [1]{\csname bibitem#1\endcsname}%
\bibitem{Kapec:2017gsg}%
  \BibitemOpen
  \bibfield{author}{%
  \bibinfo {author} {\bibfnamefont{Daniel}\ \bibnamefont{Kapec}}\ and\ \bibinfo
  {author} {\bibfnamefont{Prahar}\ \bibnamefont{Mitra}},\ }%
  \bibfield{title}{%
  \enquote{\bibinfo {title} {{A $d$-Dimensional Stress Tensor for Mink$_{d+2}$
  Gravity}},}\ }%
  \bibfield{journal}{%
  \Doi{10.1007/JHEP05(2018)186}{\bibinfo {journal} {JHEP}}\ }%
  \textbf{\bibinfo {volume} {05}},\ \bibinfo {pages} {186} (\bibinfo {year}
  {2018}),\ \Eprint{http://arxiv.org/abs/1711.04371}{arXiv:1711.04371
  [hep-th]}%
  \bibAnnoteFile{NoStop}{Kapec:2017gsg}%
\bibitem{Nande:2017dba}%
  \BibitemOpen
  \bibfield{author}{%
  \bibinfo {author} {\bibfnamefont{Anjalika}\ \bibnamefont{Nande}}, \bibinfo
  {author} {\bibfnamefont{Monica}\ \bibnamefont{Pate}},\ and\ \bibinfo {author}
  {\bibfnamefont{Andrew}\ \bibnamefont{Strominger}},\ }%
  \bibfield{title}{%
  \enquote{\bibinfo {title} {{Soft Factorization in QED from 2D Kac-Moody
  Symmetry}},}\ }%
  \bibfield{journal}{%
  \Doi{10.1007/JHEP02(2018)079}{\bibinfo {journal} {JHEP}}\ }%
  \textbf{\bibinfo {volume} {02}},\ \bibinfo {pages} {079} (\bibinfo {year}
  {2018}),\ \Eprint{http://arxiv.org/abs/1705.00608}{arXiv:1705.00608
  [hep-th]}%
  \bibAnnoteFile{NoStop}{Nande:2017dba}%
\bibitem{Himwich:2020rro}%
  \BibitemOpen
  \bibfield{author}{%
  \bibinfo {author} {\bibfnamefont{Elizabeth}\ \bibnamefont{Himwich}}, \bibinfo
  {author} {\bibfnamefont{Sruthi~A.}\ \bibnamefont{Narayanan}}, \bibinfo
  {author} {\bibfnamefont{Monica}\ \bibnamefont{Pate}}, \bibinfo {author}
  {\bibfnamefont{Nisarga}\ \bibnamefont{Paul}},\ and\ \bibinfo {author}
  {\bibfnamefont{Andrew}\ \bibnamefont{Strominger}},\ }%
  \bibfield{title}{%
  \enquote{\bibinfo {title} {{The Soft $\mathcal{S}$-Matrix in Gravity}},}\ }%
  \bibfield{journal}{%
  \Doi{10.1007/JHEP09(2020)129}{\bibinfo {journal} {JHEP}}\ }%
  \textbf{\bibinfo {volume} {09}},\ \bibinfo {pages} {129} (\bibinfo {year}
  {2020}),\ \Eprint{http://arxiv.org/abs/2005.13433}{arXiv:2005.13433
  [hep-th]}%
  \bibAnnoteFile{NoStop}{Himwich:2020rro}%
\bibitem{Arkani-Hamed:2020gyp}%
  \BibitemOpen
  \bibfield{author}{%
  \bibinfo {author} {\bibfnamefont{Nima}\ \bibnamefont{Arkani-Hamed}}, \bibinfo
  {author} {\bibfnamefont{Monica}\ \bibnamefont{Pate}}, \bibinfo {author}
  {\bibfnamefont{Ana-Maria}\ \bibnamefont{Raclariu}},\ and\ \bibinfo {author}
  {\bibfnamefont{Andrew}\ \bibnamefont{Strominger}},\ }%
  \bibfield{title}{%
  \enquote{\bibinfo {title} {{Celestial amplitudes from UV to IR}},}\ }%
  \bibfield{journal}{%
  \Doi{10.1007/JHEP08(2021)062}{\bibinfo {journal} {JHEP}}\ }%
  \textbf{\bibinfo {volume} {08}},\ \bibinfo {pages} {062} (\bibinfo {year}
  {2021}),\ \Eprint{http://arxiv.org/abs/2012.04208}{arXiv:2012.04208
  [hep-th]}%
  \bibAnnoteFile{NoStop}{Arkani-Hamed:2020gyp}%
\bibitem{Magnea:2021fvy}%
  \BibitemOpen
  \bibfield{author}{%
  \bibinfo {author} {\bibfnamefont{Lorenzo}\ \bibnamefont{Magnea}},\ }%
  \bibfield{title}{%
  \enquote{\bibinfo {title} {{Non-abelian infrared divergences on the celestial
  sphere}},}\ }%
  \bibfield{journal}{%
  \Doi{10.1007/JHEP05(2021)282}{\bibinfo {journal} {JHEP}}\ }%
  \textbf{\bibinfo {volume} {05}},\ \bibinfo {pages} {282} (\bibinfo {year}
  {2021}),\ \Eprint{http://arxiv.org/abs/2104.10254}{arXiv:2104.10254
  [hep-th]}%
  \bibAnnoteFile{NoStop}{Magnea:2021fvy}%
\bibitem{Gonzalez:2021dxw}%
  \BibitemOpen
  \bibfield{author}{%
  \bibinfo {author} {\bibfnamefont{Hern\'an~A.}\ \bibnamefont{Gonz\'alez}}\
  and\ \bibinfo {author} {\bibfnamefont{Francisco}\ \bibnamefont{Rojas}},\ }%
  \bibfield{title}{%
  \enquote{\bibinfo {title} {{The structure of IR divergences in celestial
  gluon amplitudes}},}\ }%
  \bibfield{journal}{%
  \Doi{10.1007/JHEP06(2021)171}{\bibinfo {journal} {JHEP}}\ }%
  \textbf{\bibinfo {volume} {2021}},\ \bibinfo {pages} {171} (\bibinfo {year}
  {2021}),\ \Eprint{http://arxiv.org/abs/2104.12979}{arXiv:2104.12979
  [hep-th]}%
  \bibAnnoteFile{NoStop}{Gonzalez:2021dxw}%
\bibitem{Nguyen:2021qkt}%
  \BibitemOpen
  \bibfield{author}{%
  \bibinfo {author} {\bibfnamefont{Kevin}\ \bibnamefont{Nguyen}}\ and\ \bibinfo
  {author} {\bibfnamefont{Jakob}\ \bibnamefont{Salzer}},\ }%
  \bibfield{title}{%
  \enquote{\bibinfo {title} {{Celestial IR divergences and the effective action
  of supertranslation modes}},}\ }%
   (\bibinfo {month} {5}\ \bibinfo {year} {2021}),\
  \Eprint{http://arxiv.org/abs/2105.10526}{arXiv:2105.10526 [hep-th]}%
  \bibAnnoteFile{NoStop}{Nguyen:2021qkt}%
\bibitem{Kalyanapuram:2021tnl}%
  \BibitemOpen
  \bibfield{author}{%
  \bibinfo {author} {\bibfnamefont{Nikhil}\ \bibnamefont{Kalyanapuram}},\ }%
  \bibfield{title}{%
  \enquote{\bibinfo {title} {{Infrared and Holographic Aspects of the
  $S$-Matrix in Gauge Theory and Gravity}},}\ }%
   (\bibinfo {month} {7}\ \bibinfo {year} {2021}),\
  \Eprint{http://arxiv.org/abs/2107.06660}{arXiv:2107.06660 [hep-th]}%
  \bibAnnoteFile{NoStop}{Kalyanapuram:2021tnl}%
\bibitem{He:2014laa}%
  \BibitemOpen
  \bibfield{author}{%
  \bibinfo {author} {\bibfnamefont{Temple}\ \bibnamefont{He}}, \bibinfo
  {author} {\bibfnamefont{Vyacheslav}\ \bibnamefont{Lysov}}, \bibinfo {author}
  {\bibfnamefont{Prahar}\ \bibnamefont{Mitra}},\ and\ \bibinfo {author}
  {\bibfnamefont{Andrew}\ \bibnamefont{Strominger}},\ }%
  \bibfield{title}{%
  \enquote{\bibinfo {title} {{BMS supertranslations and Weinberg's soft
  graviton theorem}},}\ }%
  \bibfield{journal}{%
  \Doi{10.1007/JHEP05(2015)151}{\bibinfo {journal} {JHEP}}\ }%
  \textbf{\bibinfo {volume} {05}},\ \bibinfo {pages} {151} (\bibinfo {year}
  {2015}),\ \Eprint{http://arxiv.org/abs/1401.7026}{arXiv:1401.7026 [hep-th]}%
  \bibAnnoteFile{NoStop}{He:2014laa}%
\bibitem{He:2014cra}%
  \BibitemOpen
  \bibfield{author}{%
  \bibinfo {author} {\bibfnamefont{Temple}\ \bibnamefont{He}}, \bibinfo
  {author} {\bibfnamefont{Prahar}\ \bibnamefont{Mitra}}, \bibinfo {author}
  {\bibfnamefont{Achilleas~P.}\ \bibnamefont{Porfyriadis}},\ and\ \bibinfo
  {author} {\bibfnamefont{Andrew}\ \bibnamefont{Strominger}},\ }%
  \bibfield{title}{%
  \enquote{\bibinfo {title} {{New Symmetries of Massless QED}},}\ }%
  \bibfield{journal}{%
  \Doi{10.1007/JHEP10(2014)112}{\bibinfo {journal} {JHEP}}\ }%
  \textbf{\bibinfo {volume} {10}},\ \bibinfo {pages} {112} (\bibinfo {year}
  {2014}),\ \Eprint{http://arxiv.org/abs/1407.3789}{arXiv:1407.3789 [hep-th]}%
  \bibAnnoteFile{NoStop}{He:2014cra}%
\bibitem{Kapec:2014opa}%
  \BibitemOpen
  \bibfield{author}{%
  \bibinfo {author} {\bibfnamefont{Daniel}\ \bibnamefont{Kapec}}, \bibinfo
  {author} {\bibfnamefont{Vyacheslav}\ \bibnamefont{Lysov}}, \bibinfo {author}
  {\bibfnamefont{Sabrina}\ \bibnamefont{Pasterski}},\ and\ \bibinfo {author}
  {\bibfnamefont{Andrew}\ \bibnamefont{Strominger}},\ }%
  \bibfield{title}{%
  \enquote{\bibinfo {title} {{Semiclassical Virasoro symmetry of the quantum
  gravity $ \mathcal{S}$-matrix}},}\ }%
  \bibfield{journal}{%
  \Doi{10.1007/JHEP08(2014)058}{\bibinfo {journal} {JHEP}}\ }%
  \textbf{\bibinfo {volume} {08}},\ \bibinfo {pages} {058} (\bibinfo {year}
  {2014}),\ \Eprint{http://arxiv.org/abs/1406.3312}{arXiv:1406.3312 [hep-th]}%
  \bibAnnoteFile{NoStop}{Kapec:2014opa}%
\bibitem{He:2015zea}%
  \BibitemOpen
  \bibfield{author}{%
  \bibinfo {author} {\bibfnamefont{Temple}\ \bibnamefont{He}}, \bibinfo
  {author} {\bibfnamefont{Prahar}\ \bibnamefont{Mitra}},\ and\ \bibinfo
  {author} {\bibfnamefont{Andrew}\ \bibnamefont{Strominger}},\ }%
  \bibfield{title}{%
  \enquote{\bibinfo {title} {{2D Kac-Moody Symmetry of 4D Yang-Mills
  Theory}},}\ }%
  \bibfield{journal}{%
  \Doi{10.1007/JHEP10(2016)137}{\bibinfo {journal} {JHEP}}\ }%
  \textbf{\bibinfo {volume} {10}},\ \bibinfo {pages} {137} (\bibinfo {year}
  {2016}),\ \Eprint{http://arxiv.org/abs/1503.02663}{arXiv:1503.02663
  [hep-th]}%
  \bibAnnoteFile{NoStop}{He:2015zea}%
\bibitem{Kapec:2016jld}%
  \BibitemOpen
  \bibfield{author}{%
  \bibinfo {author} {\bibfnamefont{Daniel}\ \bibnamefont{Kapec}}, \bibinfo
  {author} {\bibfnamefont{Prahar}\ \bibnamefont{Mitra}}, \bibinfo {author}
  {\bibfnamefont{Ana-Maria}\ \bibnamefont{Raclariu}},\ and\ \bibinfo {author}
  {\bibfnamefont{Andrew}\ \bibnamefont{Strominger}},\ }%
  \bibfield{title}{%
  \enquote{\bibinfo {title} {{2D Stress Tensor for 4D Gravity}},}\ }%
  \bibfield{journal}{%
  \Doi{10.1103/PhysRevLett.119.121601}{\bibinfo {journal} {Phys. Rev. Lett.}}\
  }%
  \textbf{\bibinfo {volume} {119}},\ \bibinfo {pages} {121601} (\bibinfo {year}
  {2017}),\ \Eprint{http://arxiv.org/abs/1609.00282}{arXiv:1609.00282
  [hep-th]}%
  \bibAnnoteFile{NoStop}{Kapec:2016jld}%
\bibitem{He:2017fsb}%
  \BibitemOpen
  \bibfield{author}{%
  \bibinfo {author} {\bibfnamefont{Temple}\ \bibnamefont{He}}, \bibinfo
  {author} {\bibfnamefont{Daniel}\ \bibnamefont{Kapec}}, \bibinfo {author}
  {\bibfnamefont{Ana-Maria}\ \bibnamefont{Raclariu}},\ and\ \bibinfo {author}
  {\bibfnamefont{Andrew}\ \bibnamefont{Strominger}},\ }%
  \bibfield{title}{%
  \enquote{\bibinfo {title} {{Loop-Corrected Virasoro Symmetry of 4D Quantum
  Gravity}},}\ }%
  \bibfield{journal}{%
  \Doi{10.1007/JHEP08(2017)050}{\bibinfo {journal} {JHEP}}\ }%
  \textbf{\bibinfo {volume} {08}},\ \bibinfo {pages} {050} (\bibinfo {year}
  {2017}),\ \Eprint{http://arxiv.org/abs/1701.00496}{arXiv:1701.00496
  [hep-th]}%
  \bibAnnoteFile{NoStop}{He:2017fsb}%
\bibitem{Weinberg:1965nx}%
  \BibitemOpen
  \bibfield{author}{%
  \bibinfo {author} {\bibfnamefont{Steven}\ \bibnamefont{Weinberg}},\ }%
  \bibfield{title}{%
  \enquote{\bibinfo {title} {{Infrared photons and gravitons}},}\ }%
  \bibfield{journal}{%
  \Doi{10.1103/PhysRev.140.B516}{\bibinfo {journal} {Phys. Rev.}}\ }%
  \textbf{\bibinfo {volume} {140}},\ \bibinfo {pages} {B516--B524} (\bibinfo
  {year} {1965})%
  \bibAnnoteFile{NoStop}{Weinberg:1965nx}%
\bibitem{Strominger:2013jfa}%
  \BibitemOpen
  \bibfield{author}{%
  \bibinfo {author} {\bibfnamefont{Andrew}\ \bibnamefont{Strominger}},\ }%
  \bibfield{title}{%
  \enquote{\bibinfo {title} {{On BMS Invariance of Gravitational
  Scattering}},}\ }%
  \bibfield{journal}{%
  \Doi{10.1007/JHEP07(2014)152}{\bibinfo {journal} {JHEP}}\ }%
  \textbf{\bibinfo {volume} {07}},\ \bibinfo {pages} {152} (\bibinfo {year}
  {2014}),\ \Eprint{http://arxiv.org/abs/1312.2229}{arXiv:1312.2229 [hep-th]}%
  \bibAnnoteFile{NoStop}{Strominger:2013jfa}%
\bibitem{mott_1931}%
  \BibitemOpen
  \bibfield{author}{%
  \bibinfo {author} {\bibfnamefont{N.~F.}\ \bibnamefont{Mott}},\ }%
  \bibfield{title}{%
  \enquote{\bibinfo {title} {On the influence of radiative forces on the
  scattering of electrons},}\ }%
  \bibfield{journal}{%
  \Doi{10.1017/S0305004100010379}{\bibinfo {journal} {Mathematical Proceedings
  of the Cambridge Philosophical Society}}\ }%
  \textbf{\bibinfo {volume} {27}},\ \bibinfo {pages} {255--267} (\bibinfo
  {year} {1931})%
  \bibAnnoteFile{NoStop}{mott_1931}%
\bibitem{Bloch:1937pw}%
  \BibitemOpen
  \bibfield{author}{%
  \bibinfo {author} {\bibfnamefont{F.}~\bibnamefont{Bloch}}\ and\ \bibinfo
  {author} {\bibfnamefont{A.}~\bibnamefont{Nordsieck}},\ }%
  \bibfield{title}{%
  \enquote{\bibinfo {title} {{Note on the Radiation Field of the electron}},}\
  }%
  \bibfield{journal}{%
  \Doi{10.1103/PhysRev.52.54}{\bibinfo {journal} {Phys. Rev.}}\ }%
  \textbf{\bibinfo {volume} {52}},\ \bibinfo {pages} {54--59} (\bibinfo {year}
  {1937})%
  \bibAnnoteFile{NoStop}{Bloch:1937pw}%
\bibitem{Chung:1965zza}%
  \BibitemOpen
  \bibfield{author}{%
  \bibinfo {author} {\bibfnamefont{Victor}\ \bibnamefont{Chung}},\ }%
  \bibfield{title}{%
  \enquote{\bibinfo {title} {{Infrared Divergence in Quantum
  Electrodynamics}},}\ }%
  \bibfield{journal}{%
  \Doi{10.1103/PhysRev.140.B1110}{\bibinfo {journal} {Phys. Rev.}}\ }%
  \textbf{\bibinfo {volume} {140}},\ \bibinfo {pages} {B1110--B1122} (\bibinfo
  {year} {1965})%
  \bibAnnoteFile{NoStop}{Chung:1965zza}%
\bibitem{Kibble:1968sfb}%
  \BibitemOpen
  \bibfield{author}{%
  \bibinfo {author} {\bibfnamefont{T.~W.~B.}\ \bibnamefont{Kibble}},\ }%
  \bibfield{title}{%
  \enquote{\bibinfo {title} {{Coherent Soft-Photon States and Infrared
  Divergences. I. Classical Currents}},}\ }%
  \bibfield{journal}{%
  \Doi{10.1063/1.1664582}{\bibinfo {journal} {J. Math. Phys.}}\ }%
  \textbf{\bibinfo {volume} {9}},\ \bibinfo {pages} {315--324} (\bibinfo {year}
  {1968})%
  \bibAnnoteFile{NoStop}{Kibble:1968sfb}%
\bibitem{Kibble:1969ip}%
  \BibitemOpen
  \bibfield{author}{%
  \bibinfo {author} {\bibfnamefont{T.~W.~B.}\ \bibnamefont{Kibble}},\ }%
  \bibfield{title}{%
  \enquote{\bibinfo {title} {{Coherent soft-photon states and infrared
  divergences. ii. mass-shell singularities of green's functions}},}\ }%
  \bibfield{journal}{%
  \Doi{10.1103/PhysRev.173.1527}{\bibinfo {journal} {Phys. Rev.}}\ }%
  \textbf{\bibinfo {volume} {173}},\ \bibinfo {pages} {1527--1535} (\bibinfo
  {year} {1968})%
  \bibAnnoteFile{NoStop}{Kibble:1969ip}%
\bibitem{Kibble:1969ep}%
  \BibitemOpen
  \bibfield{author}{%
  \bibinfo {author} {\bibfnamefont{T.~W.~B.}\ \bibnamefont{Kibble}},\ }%
  \bibfield{title}{%
  \enquote{\bibinfo {title} {{Coherent soft-photon states and infrared
  divergences. iii. asymptotic states and reduction formulas}},}\ }%
  \bibfield{journal}{%
  \Doi{10.1103/PhysRev.174.1882}{\bibinfo {journal} {Phys. Rev.}}\ }%
  \textbf{\bibinfo {volume} {174}},\ \bibinfo {pages} {1882--1901} (\bibinfo
  {year} {1968})%
  \bibAnnoteFile{NoStop}{Kibble:1969ep}%
\bibitem{Kibble:1969kd}%
  \BibitemOpen
  \bibfield{author}{%
  \bibinfo {author} {\bibfnamefont{T.~W.~B.}\ \bibnamefont{Kibble}},\ }%
  \bibfield{title}{%
  \enquote{\bibinfo {title} {{Coherent soft-photon states and infrared
  divergences. iv. the scattering operator}},}\ }%
  \bibfield{journal}{%
  \Doi{10.1103/PhysRev.175.1624}{\bibinfo {journal} {Phys. Rev.}}\ }%
  \textbf{\bibinfo {volume} {175}},\ \bibinfo {pages} {1624--1640} (\bibinfo
  {year} {1968})%
  \bibAnnoteFile{NoStop}{Kibble:1969kd}%
\bibitem{Kulish:1970ut}%
  \BibitemOpen
  \bibfield{author}{%
  \bibinfo {author} {\bibfnamefont{P.P.}\ \bibnamefont{Kulish}}\ and\ \bibinfo
  {author} {\bibfnamefont{L.D.}\ \bibnamefont{Faddeev}},\ }%
  \bibfield{title}{%
  \enquote{\bibinfo {title} {{Asymptotic conditions and infrared divergences in
  quantum electrodynamics}},}\ }%
  \bibfield{journal}{%
  \Doi{10.1007/BF01066485}{\bibinfo {journal} {Theor. Math. Phys.}}\ }%
  \textbf{\bibinfo {volume} {4}},\ \bibinfo {pages} {745} (\bibinfo {year}
  {1970})%
  \bibAnnoteFile{NoStop}{Kulish:1970ut}%
\bibitem{Kapec:2017tkm}%
  \BibitemOpen
  \bibfield{author}{%
  \bibinfo {author} {\bibfnamefont{Daniel}\ \bibnamefont{Kapec}}, \bibinfo
  {author} {\bibfnamefont{Malcolm}\ \bibnamefont{Perry}}, \bibinfo {author}
  {\bibfnamefont{Ana-Maria}\ \bibnamefont{Raclariu}},\ and\ \bibinfo {author}
  {\bibfnamefont{Andrew}\ \bibnamefont{Strominger}},\ }%
  \bibfield{title}{%
  \enquote{\bibinfo {title} {{Infrared Divergences in QED, Revisited}},}\ }%
  \bibfield{journal}{%
  \Doi{10.1103/PhysRevD.96.085002}{\bibinfo {journal} {Phys. Rev.}}\ }%
  \textbf{\bibinfo {volume} {D96}},\ \bibinfo {pages} {085002} (\bibinfo {year}
  {2017}),\ \Eprint{http://arxiv.org/abs/1705.04311}{arXiv:1705.04311
  [hep-th]}%
  \bibAnnoteFile{NoStop}{Kapec:2017tkm}%
\bibitem{Weinberg:1995mt}%
  \BibitemOpen
  \bibfield{author}{%
  \bibinfo {author} {\bibfnamefont{Steven}\ \bibnamefont{Weinberg}},\ }%
  \emph{\bibinfo {title} {{The Quantum theory of fields. Vol. 1:
  Foundations}}}\ (\bibinfo {publisher} {Cambridge University Press},\ \bibinfo
  {year} {2005})\ ISBN \bibinfo {isbn} {978-0-521-67053-1, 978-0-511-25204-4}%
  \bibAnnoteFile{NoStop}{Weinberg:1995mt}%
\bibitem{Strominger:2015bla}%
  \BibitemOpen
  \bibfield{author}{%
  \bibinfo {author} {\bibfnamefont{Andrew}\ \bibnamefont{Strominger}},\ }%
  \bibfield{title}{%
  \enquote{\bibinfo {title} {{Magnetic Corrections to the Soft Photon
  Theorem}},}\ }%
  \bibfield{journal}{%
  \Doi{10.1103/PhysRevLett.116.031602}{\bibinfo {journal} {Phys. Rev. Lett.}}\
  }%
  \textbf{\bibinfo {volume} {116}},\ \bibinfo {pages} {031602} (\bibinfo {year}
  {2016}),\ \Eprint{http://arxiv.org/abs/1509.00543}{arXiv:1509.00543
  [hep-th]}%
  \bibAnnoteFile{NoStop}{Strominger:2015bla}%
\bibitem{Goddard_1978}%
  \BibitemOpen
  \bibfield{author}{%
  \bibinfo {author} {\bibfnamefont{P}~\bibnamefont{Goddard}}\ and\ \bibinfo
  {author} {\bibfnamefont{D~I}\ \bibnamefont{Olive}},\ }%
  \bibfield{title}{%
  \enquote{\bibinfo {title} {Magnetic monopoles in gauge field theories},}\ }%
  \bibfield{journal}{%
  \Doi{10.1088/0034-4885/41/9/001}{\bibinfo {journal} {Reports on Progress in
  Physics}}\ }%
  \textbf{\bibinfo {volume} {41}},\ \bibinfo {pages} {1357--1437} (\bibinfo
  {month} {sep}\ \bibinfo {year} {1978}),\
  \url{https://doi.org/10.1088%2F0034-4885%2F41%2F9%2F001}%
  \bibAnnoteFile{NoStop}{Goddard_1978}%
\bibitem{Olive:1995sw}%
  \BibitemOpen
  \bibfield{author}{%
  \bibinfo {author} {\bibfnamefont{David~I}\ \bibnamefont{Olive}},\ }%
  \bibfield{title}{%
  \enquote{\bibinfo {title} {{Exact electromagnetic duality}},}\ }%
  \bibfield{journal}{%
  \Doi{10.1016/0920-5632(96)00002-3}{\bibinfo {journal} {Nucl. Phys. B Proc.
  Suppl.}}\ }%
  \textbf{\bibinfo {volume} {45}},\ \bibinfo {pages} {88--102} (\bibinfo {year}
  {1996}),\ \Eprint{http://arxiv.org/abs/hep-th/9508089}{arXiv:hep-th/9508089}%
  \bibAnnoteFile{NoStop}{Olive:1995sw}%
\bibitem{Weinberg:1965rz}%
  \BibitemOpen
  \bibfield{author}{%
  \bibinfo {author} {\bibfnamefont{Steven}\ \bibnamefont{Weinberg}},\ }%
  \bibfield{title}{%
  \enquote{\bibinfo {title} {{Photons and gravitons in perturbation theory:
  Derivation of Maxwell's and Einstein's equations}},}\ }%
  \bibfield{journal}{%
  \Doi{10.1103/PhysRev.138.B988}{\bibinfo {journal} {Phys. Rev.}}\ }%
  \textbf{\bibinfo {volume} {138}},\ \bibinfo {pages} {B988--B1002} (\bibinfo
  {year} {1965})%
  \bibAnnoteFile{NoStop}{Weinberg:1965rz}%
\bibitem{Terning:2018udc}%
  \BibitemOpen
  \bibfield{author}{%
  \bibinfo {author} {\bibfnamefont{John}\ \bibnamefont{Terning}}\ and\ \bibinfo
  {author} {\bibfnamefont{Christopher~B.}\ \bibnamefont{Verhaaren}},\ }%
  \bibfield{title}{%
  \enquote{\bibinfo {title} {{Resolving the Weinberg Paradox with Topology}},}\
  }%
  \bibfield{journal}{%
  \Doi{10.1007/JHEP03(2019)177}{\bibinfo {journal} {JHEP}}\ }%
  \textbf{\bibinfo {volume} {03}},\ \bibinfo {pages} {177} (\bibinfo {year}
  {2019}),\ \Eprint{http://arxiv.org/abs/1809.05102}{arXiv:1809.05102
  [hep-th]}%
  \bibAnnoteFile{NoStop}{Terning:2018udc}%
\bibitem{Goddard:1976qe}%
  \BibitemOpen
  \bibfield{author}{%
  \bibinfo {author} {\bibfnamefont{P.}~\bibnamefont{Goddard}}, \bibinfo
  {author} {\bibfnamefont{J.}~\bibnamefont{Nuyts}},\ and\ \bibinfo {author}
  {\bibfnamefont{David~I.}\ \bibnamefont{Olive}},\ }%
  \bibfield{title}{%
  \enquote{\bibinfo {title} {{Gauge Theories and Magnetic Charge}},}\ }%
  \bibfield{journal}{%
  \Doi{10.1016/0550-3213(77)90221-8}{\bibinfo {journal} {Nucl. Phys. B}}\ }%
  \textbf{\bibinfo {volume} {125}},\ \bibinfo {pages} {1--28} (\bibinfo {year}
  {1977})%
  \bibAnnoteFile{NoStop}{Goddard:1976qe}%
\bibitem{Abouelsaood:1982dz}%
  \BibitemOpen
  \bibfield{author}{%
  \bibinfo {author} {\bibfnamefont{Ahmed}\ \bibnamefont{Abouelsaood}},\ }%
  \bibfield{title}{%
  \enquote{\bibinfo {title} {{Are There Chromodyons?}}.}\ }%
  \bibfield{journal}{%
  \Doi{10.1016/0550-3213(83)90195-5}{\bibinfo {journal} {Nucl. Phys. B}}\ }%
  \textbf{\bibinfo {volume} {226}},\ \bibinfo {pages} {309--338} (\bibinfo
  {year} {1983})%
  \bibAnnoteFile{NoStop}{Abouelsaood:1982dz}%
\bibitem{Nelson:1983em}%
  \BibitemOpen
  \bibfield{author}{%
  \bibinfo {author} {\bibfnamefont{Philip~C.}\ \bibnamefont{Nelson}},\ }%
  \bibfield{title}{%
  \enquote{\bibinfo {title} {{Excitations of SU(5) Monopoles}},}\ }%
  \bibfield{journal}{%
  \Doi{10.1103/PhysRevLett.50.939}{\bibinfo {journal} {Phys. Rev. Lett.}}\ }%
  \textbf{\bibinfo {volume} {50}},\ \bibinfo {pages} {939} (\bibinfo {year}
  {1983})%
  \bibAnnoteFile{NoStop}{Nelson:1983em}%
\bibitem{Balachandran:1982gt}%
  \BibitemOpen
  \bibfield{author}{%
  \bibinfo {author} {\bibfnamefont{A.~P.}\ \bibnamefont{Balachandran}},
  \bibinfo {author} {\bibfnamefont{G.}~\bibnamefont{Marmo}}, \bibinfo {author}
  {\bibfnamefont{N.}~\bibnamefont{Mukunda}}, \bibinfo {author}
  {\bibfnamefont{J.~S.}\ \bibnamefont{Nilsson}}, \bibinfo {author}
  {\bibfnamefont{E.~C.~G.}\ \bibnamefont{Sudarshan}},\ and\ \bibinfo {author}
  {\bibfnamefont{F.}~\bibnamefont{Zaccaria}},\ }%
  \bibfield{title}{%
  \enquote{\bibinfo {title} {{Monopole Topology and the Problem of Color}},}\
  }%
  \bibfield{journal}{%
  \Doi{10.1103/PhysRevLett.50.1553}{\bibinfo {journal} {Phys. Rev. Lett.}}\ }%
  \textbf{\bibinfo {volume} {50}},\ \bibinfo {pages} {1553} (\bibinfo {year}
  {1983})%
  \bibAnnoteFile{NoStop}{Balachandran:1982gt}%
\bibitem{Nelson:1983bu}%
  \BibitemOpen
  \bibfield{author}{%
  \bibinfo {author} {\bibfnamefont{Philip~C.}\ \bibnamefont{Nelson}}\ and\
  \bibinfo {author} {\bibfnamefont{Aneesh}\ \bibnamefont{Manohar}},\ }%
  \bibfield{title}{%
  \enquote{\bibinfo {title} {{Global Color Is Not Always Defined}},}\ }%
  \bibfield{journal}{%
  \Doi{10.1103/PhysRevLett.50.943}{\bibinfo {journal} {Phys. Rev. Lett.}}\ }%
  \textbf{\bibinfo {volume} {50}},\ \bibinfo {pages} {943} (\bibinfo {year}
  {1983})%
  \bibAnnoteFile{NoStop}{Nelson:1983bu}%
\bibitem{Balachandran:1983xz}%
  \BibitemOpen
  \bibfield{author}{%
  \bibinfo {author} {\bibfnamefont{A.~P.}\ \bibnamefont{Balachandran}},
  \bibinfo {author} {\bibfnamefont{G.}~\bibnamefont{Marmo}}, \bibinfo {author}
  {\bibfnamefont{N.}~\bibnamefont{Mukunda}}, \bibinfo {author}
  {\bibfnamefont{J.~S.}\ \bibnamefont{Nilsson}}, \bibinfo {author}
  {\bibfnamefont{E.~C.~G.}\ \bibnamefont{Sudarshan}},\ and\ \bibinfo {author}
  {\bibfnamefont{F.}~\bibnamefont{Zaccaria}},\ }%
  \bibfield{title}{%
  \enquote{\bibinfo {title} {{Nonabelian Monopoles Break Color. 1. Classical
  Mechanics}},}\ }%
  \bibfield{journal}{%
  \Doi{10.1103/PhysRevD.29.2919}{\bibinfo {journal} {Phys. Rev. D}}\ }%
  \textbf{\bibinfo {volume} {29}},\ \bibinfo {pages} {2919} (\bibinfo {year}
  {1984})%
  \bibAnnoteFile{NoStop}{Balachandran:1983xz}%
\bibitem{Balachandran:1983fg}%
  \BibitemOpen
  \bibfield{author}{%
  \bibinfo {author} {\bibfnamefont{A.~P.}\ \bibnamefont{Balachandran}},
  \bibinfo {author} {\bibfnamefont{G.}~\bibnamefont{Marmo}}, \bibinfo {author}
  {\bibfnamefont{N.}~\bibnamefont{Mukunda}}, \bibinfo {author}
  {\bibfnamefont{J.~S.}\ \bibnamefont{Nilsson}}, \bibinfo {author}
  {\bibfnamefont{E.~C.~G.}\ \bibnamefont{Sudarshan}},\ and\ \bibinfo {author}
  {\bibfnamefont{F.}~\bibnamefont{Zaccaria}},\ }%
  \bibfield{title}{%
  \enquote{\bibinfo {title} {{Nonabelian Monopoles Break Color. 2. Field Theory
  and Quantum Mechanics}},}\ }%
  \bibfield{journal}{%
  \Doi{10.1103/PhysRevD.29.2936}{\bibinfo {journal} {Phys. Rev. D}}\ }%
  \textbf{\bibinfo {volume} {29}},\ \bibinfo {pages} {2936} (\bibinfo {year}
  {1984})%
  \bibAnnoteFile{NoStop}{Balachandran:1983fg}%
\bibitem{Nelson:1983fn}%
  \BibitemOpen
  \bibfield{author}{%
  \bibinfo {author} {\bibfnamefont{Philip~C.}\ \bibnamefont{Nelson}}\ and\
  \bibinfo {author} {\bibfnamefont{Sidney~R.}\ \bibnamefont{Coleman}},\ }%
  \bibfield{title}{%
  \enquote{\bibinfo {title} {{What Becomes of Global Color}},}\ }%
  \bibfield{journal}{%
  \Doi{10.1016/0550-3213(84)90013-0}{\bibinfo {journal} {Nucl. Phys. B}}\ }%
  \textbf{\bibinfo {volume} {237}},\ \bibinfo {pages} {1--31} (\bibinfo {year}
  {1984})%
  \bibAnnoteFile{NoStop}{Nelson:1983fn}%
\bibitem{Ortin:2015hya}%
  \BibitemOpen
  \bibfield{author}{%
  \bibinfo {author} {\bibfnamefont{Tomas}\ \bibnamefont{Ortin}},\ }%
  \Doi{10.1017/CBO9781139019750}{\emph{\bibinfo {title} {{Gravity and
  Strings}}}},\ \bibinfo {edition} {2nd}\ ed.,\ Cambridge Monographs on
  Mathematical Physics\ (\bibinfo {publisher} {Cambridge University Press},\
  \bibinfo {year} {2015})\ ISBN \bibinfo {isbn} {978-0-521-76813-9,
  978-0-521-76813-9, 978-1-316-23579-9}%
  \bibAnnoteFile{NoStop}{Ortin:2015hya}%
\bibitem{Strominger:2017zoo}%
  \BibitemOpen
  \bibfield{author}{%
  \bibinfo {author} {\bibfnamefont{Andrew}\ \bibnamefont{Strominger}},\ }%
  \bibfield{title}{%
  \enquote{\bibinfo {title} {{Lectures on the Infrared Structure of Gravity and
  Gauge Theory}},}\ }%
   (\bibinfo {month} {3}\ \bibinfo {year} {2017}),\
  \Eprint{http://arxiv.org/abs/1703.05448}{arXiv:1703.05448 [hep-th]}%
  \bibAnnoteFile{NoStop}{Strominger:2017zoo}%
\bibitem{He:2019jjk}%
  \BibitemOpen
  \bibfield{author}{%
  \bibinfo {author} {\bibfnamefont{Temple}\ \bibnamefont{He}}\ and\ \bibinfo
  {author} {\bibfnamefont{Prahar}\ \bibnamefont{Mitra}},\ }%
  \bibfield{title}{%
  \enquote{\bibinfo {title} {{Asymptotic symmetries and
  Weinberg\textquoteright{}s soft photon theorem in Mink$_{d+2}$}},}\ }%
  \bibfield{journal}{%
  \Doi{10.1007/JHEP10(2019)213}{\bibinfo {journal} {JHEP}}\ }%
  \textbf{\bibinfo {volume} {10}},\ \bibinfo {pages} {213} (\bibinfo {year}
  {2019}),\ \Eprint{http://arxiv.org/abs/1903.02608}{arXiv:1903.02608
  [hep-th]}%
  \bibAnnoteFile{NoStop}{He:2019jjk}%
\bibitem{Aurilia:1993qi}%
  \BibitemOpen
  \bibfield{author}{%
  \bibinfo {author} {\bibfnamefont{Antonio}\ \bibnamefont{Aurilia}}, \bibinfo
  {author} {\bibfnamefont{Anais}\ \bibnamefont{Smailagic}},\ and\ \bibinfo
  {author} {\bibfnamefont{Euro}\ \bibnamefont{Spallucci}},\ }%
  \bibfield{title}{%
  \enquote{\bibinfo {title} {{Gauge theory of the string geodesic field}},}\ }%
  \bibfield{journal}{%
  \Doi{10.1103/PhysRevD.47.2536}{\bibinfo {journal} {Phys. Rev. D}}\ }%
  \textbf{\bibinfo {volume} {47}},\ \bibinfo {pages} {2536--2548} (\bibinfo
  {year} {1993}),\
  \Eprint{http://arxiv.org/abs/hep-th/9301019}{arXiv:hep-th/9301019}%
  \bibAnnoteFile{NoStop}{Aurilia:1993qi}%
\bibitem{Afshar:2018apx}%
  \BibitemOpen
  \bibfield{author}{%
  \bibinfo {author} {\bibfnamefont{Hamid}\ \bibnamefont{Afshar}}, \bibinfo
  {author} {\bibfnamefont{Erfan}\ \bibnamefont{Esmaeili}},\ and\ \bibinfo
  {author} {\bibfnamefont{M.~M.}\ \bibnamefont{Sheikh-Jabbari}},\ }%
  \bibfield{title}{%
  \enquote{\bibinfo {title} {{Asymptotic Symmetries in $p$-Form Theories}},}\
  }%
  \bibfield{journal}{%
  \Doi{10.1007/JHEP05(2018)042}{\bibinfo {journal} {JHEP}}\ }%
  \textbf{\bibinfo {volume} {05}},\ \bibinfo {pages} {042} (\bibinfo {year}
  {2018}),\ \Eprint{http://arxiv.org/abs/1801.07752}{arXiv:1801.07752
  [hep-th]}%
  \bibAnnoteFile{NoStop}{Afshar:2018apx}%
\bibitem{He:2019ywq}%
  \BibitemOpen
  \bibfield{author}{%
  \bibinfo {author} {\bibfnamefont{Temple}\ \bibnamefont{He}}\ and\ \bibinfo
  {author} {\bibfnamefont{Prahar}\ \bibnamefont{Mitra}},\ }%
  \bibfield{title}{%
  \enquote{\bibinfo {title} {{New magnetic symmetries in $(d + 2)$-dimensional
  QED}},}\ }%
  \bibfield{journal}{%
  \Doi{10.1007/JHEP01(2021)122}{\bibinfo {journal} {JHEP}}\ }%
  \textbf{\bibinfo {volume} {01}},\ \bibinfo {pages} {122} (\bibinfo {year}
  {2021}),\ \Eprint{http://arxiv.org/abs/1907.02808}{arXiv:1907.02808
  [hep-th]}%
  \bibAnnoteFile{NoStop}{He:2019ywq}%
\bibitem{Maldacena:2016upp}%
  \BibitemOpen
  \bibfield{author}{%
  \bibinfo {author} {\bibfnamefont{Juan}\ \bibnamefont{Maldacena}}, \bibinfo
  {author} {\bibfnamefont{Douglas}\ \bibnamefont{Stanford}},\ and\ \bibinfo
  {author} {\bibfnamefont{Zhenbin}\ \bibnamefont{Yang}},\ }%
  \bibfield{title}{%
  \enquote{\bibinfo {title} {{Conformal symmetry and its breaking in two
  dimensional Nearly Anti-de-Sitter space}},}\ }%
  \bibfield{journal}{%
  \Doi{10.1093/ptep/ptw124}{\bibinfo {journal} {PTEP}}\ }%
  \textbf{\bibinfo {volume} {2016}},\ \bibinfo {pages} {12C104} (\bibinfo
  {year} {2016}),\ \Eprint{http://arxiv.org/abs/1606.01857}{arXiv:1606.01857
  [hep-th]}%
  \bibAnnoteFile{NoStop}{Maldacena:2016upp}%
\bibitem{Campiglia:2015qka}%
  \BibitemOpen
  \bibfield{author}{%
  \bibinfo {author} {\bibfnamefont{Miguel}\ \bibnamefont{Campiglia}}\ and\
  \bibinfo {author} {\bibfnamefont{Alok}\ \bibnamefont{Laddha}},\ }%
  \bibfield{title}{%
  \enquote{\bibinfo {title} {{Asymptotic symmetries of QED and Weinberg's soft
  photon theorem}},}\ }%
  \bibfield{journal}{%
  \Doi{10.1007/JHEP07(2015)115}{\bibinfo {journal} {JHEP}}\ }%
  \textbf{\bibinfo {volume} {07}},\ \bibinfo {pages} {115} (\bibinfo {year}
  {2015}),\ \Eprint{http://arxiv.org/abs/1505.05346}{arXiv:1505.05346
  [hep-th]}%
  \bibAnnoteFile{NoStop}{Campiglia:2015qka}%
\bibitem{Kapec:2014zla}%
  \BibitemOpen
  \bibfield{author}{%
  \bibinfo {author} {\bibfnamefont{Daniel}\ \bibnamefont{Kapec}}, \bibinfo
  {author} {\bibfnamefont{Vyacheslav}\ \bibnamefont{Lysov}},\ and\ \bibinfo
  {author} {\bibfnamefont{Andrew}\ \bibnamefont{Strominger}},\ }%
  \bibfield{title}{%
  \enquote{\bibinfo {title} {{Asymptotic Symmetries of Massless QED in Even
  Dimensions}},}\ }%
  \bibfield{journal}{%
  \Doi{10.4310/ATMP.2017.v21.n7.a6}{\bibinfo {journal} {Adv. Theor. Math.
  Phys.}}\ }%
  \textbf{\bibinfo {volume} {21}},\ \bibinfo {pages} {1747--1767} (\bibinfo
  {year} {2017}),\ \Eprint{http://arxiv.org/abs/1412.2763}{arXiv:1412.2763
  [hep-th]}%
  \bibAnnoteFile{NoStop}{Kapec:2014zla}%
\bibitem{Kapec:2015ena}%
  \BibitemOpen
  \bibfield{author}{%
  \bibinfo {author} {\bibfnamefont{Daniel}\ \bibnamefont{Kapec}}, \bibinfo
  {author} {\bibfnamefont{Monica}\ \bibnamefont{Pate}},\ and\ \bibinfo {author}
  {\bibfnamefont{Andrew}\ \bibnamefont{Strominger}},\ }%
  \bibfield{title}{%
  \enquote{\bibinfo {title} {{New Symmetries of QED}},}\ }%
  \bibfield{journal}{%
  \Doi{10.4310/ATMP.2017.v21.n7.a7}{\bibinfo {journal} {Adv. Theor. Math.
  Phys.}}\ }%
  \textbf{\bibinfo {volume} {21}},\ \bibinfo {pages} {1769--1785} (\bibinfo
  {year} {2017}),\ \Eprint{http://arxiv.org/abs/1506.02906}{arXiv:1506.02906
  [hep-th]}%
  \bibAnnoteFile{NoStop}{Kapec:2015ena}%
\bibitem{Campiglia:2018dyi}%
  \BibitemOpen
  \bibfield{author}{%
  \bibinfo {author} {\bibfnamefont{Miguel}\ \bibnamefont{Campiglia}}\ and\
  \bibinfo {author} {\bibfnamefont{Alok}\ \bibnamefont{Laddha}},\ }%
  \bibfield{title}{%
  \enquote{\bibinfo {title} {{Asymptotic charges in massless QED revisited: A
  view from Spatial Infinity}},}\ }%
  \bibfield{journal}{%
  \Doi{10.1007/JHEP05(2019)207}{\bibinfo {journal} {JHEP}}\ }%
  \textbf{\bibinfo {volume} {05}},\ \bibinfo {pages} {207} (\bibinfo {year}
  {2019}),\ \Eprint{http://arxiv.org/abs/1810.04619}{arXiv:1810.04619
  [hep-th]}%
  \bibAnnoteFile{NoStop}{Campiglia:2018dyi}%
\bibitem{Bondi:1962px}%
  \BibitemOpen
  \bibfield{author}{%
  \bibinfo {author} {\bibfnamefont{H.}~\bibnamefont{Bondi}}, \bibinfo {author}
  {\bibfnamefont{M.~G.~J.}\ \bibnamefont{van~der Burg}},\ and\ \bibinfo
  {author} {\bibfnamefont{A.~W.~K.}\ \bibnamefont{Metzner}},\ }%
  \bibfield{title}{%
  \enquote{\bibinfo {title} {{Gravitational waves in general relativity. 7.
  Waves from axisymmetric isolated systems}},}\ }%
  \bibfield{journal}{%
  \Doi{10.1098/rspa.1962.0161}{\bibinfo {journal} {Proc. Roy. Soc. Lond.}}\ }%
  \textbf{\bibinfo {volume} {A269}},\ \bibinfo {pages} {21--52} (\bibinfo
  {year} {1962})%
  \bibAnnoteFile{NoStop}{Bondi:1962px}%
\bibitem{Sachs:1962wk}%
  \BibitemOpen
  \bibfield{author}{%
  \bibinfo {author} {\bibfnamefont{R.~K.}\ \bibnamefont{Sachs}},\ }%
  \bibfield{title}{%
  \enquote{\bibinfo {title} {{Gravitational waves in general relativity. 8.
  Waves in asymptotically flat space-times}},}\ }%
  \bibfield{journal}{%
  \Doi{10.1098/rspa.1962.0206}{\bibinfo {journal} {Proc. Roy. Soc. Lond.}}\ }%
  \textbf{\bibinfo {volume} {A270}},\ \bibinfo {pages} {103--126} (\bibinfo
  {year} {1962})%
  \bibAnnoteFile{NoStop}{Sachs:1962wk}%
\bibitem{Kapec:2015vwa}%
  \BibitemOpen
  \bibfield{author}{%
  \bibinfo {author} {\bibfnamefont{Daniel}\ \bibnamefont{Kapec}}, \bibinfo
  {author} {\bibfnamefont{Vyacheslav}\ \bibnamefont{Lysov}}, \bibinfo {author}
  {\bibfnamefont{Sabrina}\ \bibnamefont{Pasterski}},\ and\ \bibinfo {author}
  {\bibfnamefont{Andrew}\ \bibnamefont{Strominger}},\ }%
  \bibfield{title}{%
  \enquote{\bibinfo {title} {{Higher-dimensional supertranslations and
  Weinberg\textquoteright{}s soft graviton theorem}},}\ }%
  \bibfield{journal}{%
  \Doi{10.4310/AMSA.2017.v2.n1.a2}{\bibinfo {journal} {Ann. Math. Sci. Appl.}}\
  }%
  \textbf{\bibinfo {volume} {02}},\ \bibinfo {pages} {69--94} (\bibinfo {year}
  {2017}),\ \Eprint{http://arxiv.org/abs/1502.07644}{arXiv:1502.07644 [gr-qc]}%
  \bibAnnoteFile{NoStop}{Kapec:2015vwa}%
\bibitem{Campiglia:2015kxa}%
  \BibitemOpen
  \bibfield{author}{%
  \bibinfo {author} {\bibfnamefont{Miguel}\ \bibnamefont{Campiglia}}\ and\
  \bibinfo {author} {\bibfnamefont{Alok}\ \bibnamefont{Laddha}},\ }%
  \bibfield{title}{%
  \enquote{\bibinfo {title} {{Asymptotic symmetries of gravity and soft
  theorems for massive particles}},}\ }%
  \bibfield{journal}{%
  \Doi{10.1007/JHEP12(2015)094}{\bibinfo {journal} {JHEP}}\ }%
  \textbf{\bibinfo {volume} {12}},\ \bibinfo {pages} {094} (\bibinfo {year}
  {2015}),\ \Eprint{http://arxiv.org/abs/1509.01406}{arXiv:1509.01406
  [hep-th]}%
  \bibAnnoteFile{NoStop}{Campiglia:2015kxa}%
\end{thebibliography}%
\end{document}